\documentclass[reprint,showpacs,preprintnumbers,amsmath,amssymb,prl,floatfix]{revtex4-1}
\usepackage[utf8]{inputenc}
\usepackage[T1]{fontenc}
\usepackage{bm}
\usepackage {amsmath, amssymb, amsfonts}  
\usepackage {graphicx}
\usepackage{xcolor}
\usepackage{physics}
\usepackage{caption}
\usepackage{subcaption}

\begin{document}

\title{Hybrid Quantum-Classical Algorithm for Hydrodynamics}

\author{Julien Zylberman$^1$}
\author{Giuseppe Di Molfetta$^2$}
\author{Marc Brachet$^3$}
\author{Nuno F. Loureiro$^4$}
\author{Fabrice Debbasch$^1$}
\affiliation{$^1$Sorbonne Université, Observatoire de Paris, Université PSL, CNRS, LERMA, F-75005 Paris, France}
\affiliation{$^2$CNRS, LIS, Aix-Marseille Université, Université de Toulon, Marseille, France}
\affiliation{$^3$ Laboratoire de Physique de l’École Normale Supérieure, ENS, Université PSL, CNRS, Sorbonne Université, Université de Paris, F-75005 Paris, France}
\affiliation{$^4$Plasma Science and Fusion Center, Massachusetts Institute of Technology, Cambridge, Massachusetts 02139, USA}

\begin{abstract}
A new model of nonlinear charged quantum relativistic fluids is presented. This model can be discretized into Discrete Time Quantum Walks (DTQWs), and a new hybrid (quantum-classical) algorithm for implementing these walks on NISQ devices is proposed. High resolution (up to $N=2^{17}$ grid points) hybrid numerical simulations of relativistic and non-relativistic hydrodynamical shocks on current IBM NISQs are performed with this algorithm and shown to reproduce equivalent simulations on classical computers. 
This work demonstrates that nonlinear fluid dynamics can be simulated on NISQs, and opens the door to simulating other, quantum and non-quantum fluids, including plasmas, with more general quantum walks and quantum automata.

\end{abstract}

\flushbottom
\maketitle

\thispagestyle{empty}

\noindent \textbf{Key points:} 
\begin{itemize}
    
\item Specific Discrete Time Quantum Walks (DTQWs) make it possible to simulate the dynamics of the classical Dirac field coupled to electromagnetic fields

\item  A generalized Madelung transformation is introduced to map the classical Dirac dynamics into potential flows of nonlinear, relativistic, quantum, charged fluids with spin immersed in electromagnetic fields. The Dirac dynamics thus connects quantum walks to hydrodynamics

\item A new hybrid, quantum-classical algorithm for DTQW-based hydrodynamical simulations is presented and illustrated by simulations of hydrodynamical shocks in the presence of electric fields.

 \item Simulations are run successfully on classical computers and on the current IBM's NISQ quantum processors, with up to $2^{17}$ grid points.

\end{itemize}

\section*{Introduction}

The so-called second quantum revolution is possibly one of the greatest scientific and technological challenges of the 21st century. One of the cornerstones of that revolution is quantum computing, {\sl i.e.}, the possibility of using quantum properties of matter to outperform current classical computers at least for several, if not all, standard computations. Quantum simulation originated with Richard Feynman \cite{Feynman1982}, who suggested using quantum systems to simulate efficiently other, more complex, quantum, and possibly also classical, systems.

Simulating efficiently the dynamics of both classical and quantum fluids is a long-standing problem in applied mathematics, and the applications in engineering and fundamental science cannot be over-estimated. It is therefore not surprising that the possible quantum simulation of fluid and plasma dynamics has already attracted considerable attention \cite{gaitan2020finding, budinski2021quantum, steijl2018parallel, steijl2019quantum, steijl2020quantum, lloyd2020quantum,Liue2026805118, engel2019quantum, doi:10.1063/5.0056974}. The aim of this article is to present a novel manner of simulating both relativistic and non-relativistic quantum fluids on existing and future quantum computers.

The Dirac equation \cite{Dirac} plays a pivotal role in this new approach. On one hand, the Dirac equation can be mapped into relativistic hydrodynamics by a generalization of the so-called Madelung transformation initially developed for the Schrödinger equation \cite{Madelung1926,Madelung1927} and later extended to the Klein-Gordon equation \cite{Wong,DEBBASCH1995255,Debbasch1997NonlinearAI} and quaternionic quantum mechanics \cite{articleLove}. On the other hand, quantum walks, which can be viewed as a quantum generalization of classical random walks \cite{Aharonov}, are a universal quantum primitive \cite{Childs}; every quantum algorithm can be expressed as a quantum walk, and several quantum walks, usually called Dirac quantum walks, admit the Dirac equation as continuous limit \cite{DIMOLFETTA2014157}. The Dirac equation can therefore be used as a bridge connecting quantum fluid dynamics to quantum walks and, thus, to quantum simulation and quantum computing.

To make the presentation definite and to keep it as simple as possible, we restrict ourselves to fluids moving in (1 + 1) dimensional space-time. But, having future applications to extreme, {\sl i.e.}, both relativistic and quantum plasmas in mind, we allow the fluid to be charged and experience an imposed constant electric field. We therefore introduce the generalization of the Madelung transformation which maps the {\sl charged} Dirac equation unto the hydrodynamics of a {\sl charged} relativistic quantum fluid, focusing on the conserved quantities, {\sl i.e.}, charge and energy-momentum.

Simulating the dynamics of this fluid through quantum walks can be done in three different ways. The first possibility is to use a classical, non-quantum computer to follow the evolution of the quantum walks and, thus, of the fluid. Alternatively, one may use a quantum computer to program the quantum walks; this approach, however, is not viable because today's 
quantum computers, also known as NISQs, are not error-free, and error accumulation would ruin the simulation much before the solution can be obtained. The third possibility is to use a hybrid quantum-classical algorithm which realises as much of the computation as possible on the NISQs available today, thus producing a hybrid simulation of a relativistic quantum fluid.

In this paper we first present the fluid equations obtained through the Madelung transformation of the Dirac equation in the relativistic and non-relativistic limit. We then present the hybrid algorithm and use it on both classical computers and NISQs to simulate shocks in a quantum fluid under the influence of an externally imposed electric field. The final section sums up our results and discusses possible extensions to other fluids, both classical and quantum, with possible coupling to arbitrary Yang-Mills and gravitational fields. Applications include in particular electromagnetic and quark-gluon plasma dynamics, both for Earth-based and astrophysical problems. The general conclusion of this work is that quantum walks can be used to simulate non-linear hydrodynamics on NISQs and future quantum computers.

\section*{Results I: Theoretical Framework}

\subsection*{Charged Dirac fluid}

It is well known that the Schr\"odinger equation can be cast into an hydrodynamic form through the so-called Madelung transformation~\cite{Madelung1926,Madelung1927}.  The Dirac equation admits a charge current and a stress-energy tensor, as all charged fluids do. The Madelung transformation for the Dirac equation is best obtained by rewriting the Dirac charge current and stress-energy tensor in terms of standard fluid variables. The Madelung transformation for the $(1 + 1)$D Dirac equation without electric field has been presented in \cite{Hatifi2019}. We now demonstrate how those results can be extended to situations where the charged $(1 + 1)$D Dirac field is coupled to a non-vanishing electric field. 

\paragraph{Dirac equation.}

In $(1 + 1)D$ flat space-time, the Dirac equation obeyed by the two component wave-function $\psi=(\psi^L,\psi^R)^T$ of a spin $1/2$ field can be written in the form
\begin{equation}
    (i\gamma^{0}D_{0}+i\gamma^{1}D_{1})\psi-m\psi=0,
\label{Dirac_equation}
\end{equation}
where $D_{0}=\partial_{t}+iqA_{0}$, $D_{1}=\partial_{x}+iqA_{1}$  
and $\gamma^0=\sigma_X=\begin{pmatrix}
    0 & 1  \\
    1 & 0,
    \end{pmatrix}$, $\gamma^1=i\sigma_Y=\begin{pmatrix}
    0 & 1  \\
    -1 & 0
    \end{pmatrix}$.
  The mass of the field is $m$, its charge is $q$, and  $(A_0, A_1)$ are the two components (in units $c=1$, $ \hbar =1$) of the vector potential acting on the field. 
Since we are working in $(1 + 1)D$ space-time, there is no magnetic field and the electric field is simply $E = -\partial_x A_0+\partial_tA_1$.

\paragraph{Charge current.}

The expressions for $D_0$ and $D_1$ entering the Dirac equation above make clear that, geometrically speaking, the potential $A_{\mu}$, with $\mu = {0, 1}$, is a connection ensuring the invariance of the Dirac equation under arbitrary local phase translations. More precisely, equation (\ref{Dirac_equation}) is invariant under the transformation $\psi(t, x) \rightarrow \exp( i q \alpha) \psi(t, x)$, $A_0 (t, x) 
 \rightarrow A_0(t, x) - \partial_t \alpha$, and $A_1 (t, x)  \rightarrow A_1(t, x) - \partial_x \alpha$, where $\alpha(t,x)$ is an arbitrary function of time and space. This invariance implies, through Noether's theorem, the conservation equation for the charge current $J$ with components $J^0 = q\bar{\psi}\gamma^0\psi$ and $J^1 = q\bar{\psi}\gamma^1\psi$, where $\bar{\psi}=\psi^{\dagger}\gamma^0$, which reads
 \begin{equation}
    \partial_tJ^0+\partial_xJ^1=0.
\end{equation}

According to standard relativistic hydrodynamics, the charge current $J$ can be expressed in terms of the scalar density $n$ and the $2$-velocity of the fluid by the 
simple relation $J = q n u$ or, equivalently, $n u = J/q = j$. Since $u$ is normalized to unity, this relation translates into $n = (j.j)^{1/2}$ and $ u = j/(j.j)^{1/2}$ where a dot 
denotes the Minkovski scalar product. In an arbitrary reference frame, the current $j$ decomposes into the fluid density $\rho = j^0$ in that frame, and into the spatial current density $\rho v = j^1$ in the same frame. The density $\rho$ in the proper frame of the space-time grid on which the walk is defined thus coincides, as it should, with $|\psi^L|^2 
+ |\psi^R|^2$. Note that $\rho$ coincides with $n$ in the local proper frame of the fluid/Dirac field.

\paragraph{Energy-momentum.}

The energy-momentum distribution of the $(1 + 1)D$ Dirac field in the presence of the electromagnetic field $A$ is described by its stress-energy tensor $T$, which reads
$T^{\mu\nu}=\frac{i}{4}(\bar{\psi}\gamma^{\mu}\partial^{\nu}\psi-\partial^{\nu}\bar{\psi}\gamma^{\mu}\psi)-\frac{1}{2}A^{\mu}J^{\nu}+(\mu \leftrightarrow \nu)$ where $J$ is the
conserved charge current. The stress-energy tensor $T$ obeys 
\begin{equation}
    \partial_{\mu}T^{\mu \nu}=F^{\nu}_{\text{ } \text{ }\mu}J^{\mu},
\end{equation}
where  $F^{\nu}_{\text{ } \text{ }\mu}=\partial^{\nu}A_{\mu}-\partial_{\mu}A^{\nu}$ is the electromagnetic tensor.  The energy-momentum of the Dirac field is {\sl not} conserved because the fluid experiences the force created by the electromagnetic field, and $F^{\nu}_{\text{ } \text{ }\mu}J^{\mu}$ is indeed the density of the Lorentz 2-force. In particular $F^{1}_{\text{ } \text{ }\mu}J^{\mu} = q\rho E$ represents the density of the electric force exerted by the electric field on the Dirac field, and $F^{0}_{\text{ } \text{ }\mu}J^{\mu}$ represents the power density of this force. 

The other main thermodynamical variable entering the macroscopic description of a relativistic fluid is the scalar enthalpy density $w$. Identifying $w$ in terms of wave-function variables is not straightforward. The density $w$ makes the contribution $w u^\mu u^\nu$ to the stress-energy tensor $T^{\mu \nu}$ of a perfect fluid. Considering the stress-energy tensor of the Dirac field leads to the identification $w = m n \cos (\phi_-)$ where $\phi_- = \phi_L - \phi_R$ is the difference between the phases of $\psi^L$ and $\psi^R$. Using the Dirac equation, the stress energy tensor can then be written as
\begin{eqnarray}
    &T^{\mu\nu}&= wu^{\mu}u^{\nu}\\
    &+&\frac{n}{4}\left[ (u^{\mu}\epsilon^{\nu\alpha}+u^{\nu}\epsilon^{\mu\alpha})\partial_{\alpha}\phi_-+(\epsilon^{\mu\alpha}\partial^{\nu}\phi_-+\epsilon^{\nu\alpha}\partial^{\mu}\phi_-)u_{\alpha} \right], \nonumber
\end{eqnarray}
where $\epsilon^{\mu \nu}$ is the Levi-Civita completely antisymmetric tensor of rank two, with the convention $\epsilon^{01} = + 1$.
The first contribution on the right-hand side is standard for relativistic perfect fluids. The other ones involve derivatives of $\phi_-$. Because of the relation between $\phi_-$ and the enthalpy per particle $w/n$, one can write
\begin{equation}
d\phi_- = \sigma \frac{1}{m} \left(1 - \frac{w}{mn}\right)^{-1/2} d\left(\frac{w}{n}\right),
\end{equation}
where $\sigma$ is the sign of $\phi_-$.
Thus, derivatives of $\phi_-$ can be rewritten as derivatives of the enthalpy per particle and all the terms which follow the perfect fluid part $w u^{\mu} u^{\nu}$ in the expression of the stress-energy tensor are therefore generalized `quantum pressure' terms, whose appearance is expected in the description of quantum fluids \cite{Madelung1926,Madelung1927,Donnelly}. 

\paragraph{Equations of motion.}

The Dirac equation can be transcribed in terms of the hydrodynamical variables.  One obtains:
\begin{equation}
    \partial_{\mu}(q nu^{\mu})=0,
\label{1}
\end{equation}
\begin{equation}
   \frac{w}{n}u^{\mu}=-\frac{1}{2}\left(\partial^{\mu}\phi_+ + \sigma \epsilon^{\mu\nu}  \frac{1}{m} (1 - \frac{w}{mn})^{-1/2} \partial_{\nu}(\frac{w}{n})\right)-qA^{\mu},
\label{abs2}
\end{equation}
\begin{equation}
    \epsilon^{\mu}_{\text{  }\text{ } \alpha}\partial_{\mu}(nu^{\alpha})=2mn\sin(\phi_-),
\label{2}
\end{equation}
where $\phi_+ = \phi_L + \phi_R$.  
The first equation is the continuity equation expressing charge conservation. The second equation is a generalization of the standard definition of potential flows for relativistic charged fluids in the presence of an electromagnetic potential $A$. The phase $\phi_+/2$ plays the role of the standard relativistic velocity potential, but there is an extra term involving the derivatives of $\phi_-$, which can be expressed in terms of $w/n$ and which actually prevents the flow from being potential. The last equation has no easy interpretation but is needed to form a set of four independent equations for the four independent hydrodynamical variables $n$, $u^1$ (related to $u^0$ via $u^0 = \sqrt {1 + (u^1)^2}$, $w$ and the potential $\phi_+$.

\subsection*{Non-relativistic flows}

In this section the Planck constant and the velocity of light are not equal to unity, i.e., $\hbar \neq 1$, $c \neq 1$, in order to see more clearly the quantum and relativistic part of the hydrodynamic equations.

The non-relativistic limit corresponds to a situation where the velocity $v$ of the fluid is much smaller than the velocity of light $c$, implying that the energy of the particle is almost equal to the rest mass energy: $E=E'+mc^2$ with $E'\ll mc^2$.
The relativistic part of the wave function has to be extracted by writing $\frac{\phi_+}{2}=\phi-mc^2t$ where we will see that $\phi$ is the non-relativistic velocity potential. More details on the limiting procedure can be found in the Supplementary Information. In the non-relativistic regime, the two-components of the wave function become identical and
the (1+1)D Dirac equation degenerates into a single, one-component Schr\"odinger equation. Then the relativistic fluid variables and equations defined in the previous section become the usual Madelung transformation of the Schrödinger equation in the presence of electromagnetic fields. The fluid density becomes $n=2r^2$ with $r=|\psi^L|=|\psi^R|$, while the fluid velocity $u^1$ becomes the usual generalized velocity $u^1=v=\frac{1}{m}(\partial_x\phi+qA_1)$.
Then the set of four independent relativistic fluid equations  (\ref{1},\ref{abs2},\ref{2}) degenerate into a set of two independent fluid equations: one expressing the conservation of matter (or charge), and another the generalization of Bernoulli equation for a potential fluid in an electromagnetic potential $V=cA_0$ and a quantum (Bohm) potential $Q=-\frac{\hbar^2}{2m}\frac{1}{\sqrt{n}}\frac{\partial^2\sqrt{n}}{\partial x^2}$, (which vanishes in the classical limit $\hbar \rightarrow 0$): 
\begin{equation}
    \partial_t n+\partial_x(nv)=0,
\end{equation}
\begin{equation}
    \partial_t \phi+\frac{1}{2}mv^2+qV+Q=0.
\end{equation}

The gradient of this Bernoulli equation leads to the inviscid Burgers' equation for a charged fluid in an electric field $E=-\partial_xV+\partial_tA_1$ and a quantum pressure force $F_Q=-\partial_xQ$:
\begin{equation}
    m\left(\partial_tv+v\partial_xv\right)=qE+F_Q.
\end{equation}

\section*{Results II: Hybrid algorithm to simulate Dirac flows}

\subsection{The hybrid algorithm}

We now present a hybrid quantum-classical algorithm based on a Discrete Time Quantum Walk (DTQW) discretisation of the charged Dirac fluid. Details on this discretisation and on the notation can be found in the Methods section.

In recent years, several circuit-based implementation schemes for DTQW have been devised and experimentally realised. The most recent implementation has been made on a five qubit trapped-ion quantum processor \cite{alderete2020quantum}. In most cases, DTQWs are implemented by blocks of multi-controlled Toffoli gates, typically of size $O(n^3)$ and depth $O(n^2)$~\cite{saeedi2013linear}, where $n$ is the number of qubits. Quite interesting is the recent scheme proposed by Asif Shakeel \cite{shakeel2020efficient}, where the basic QWs are formulated in terms of a simple Quantum-Fourier-Transform (QFT)-based circuit \cite{QFT}, polynomially improving the previous results in terms of complexity. Indeed, it yields a highly efficient and scalable, quadratic size, linear depth circuit for the basic DTQW. This makes the scheme particularly relevant to NISQ devices, on which fewer computations not only imply faster execution (as on all devices), but also reduced effects of noise and decoherence. However, current quantum-processor performances are such that the fidelity of the results decays sharply with the number of qubits necessary to perform a given task. In the specific case of the Quantum Walk, the efficiency of the implementation of the walker depends mainly on the implementation of the shift operator, since this is a multi-qubits-controlled gate. The coin operator is a one qubit gate and is easy to implement in practice, with sufficiently high fidelity. Indeed, one-qubit state operations have demonstrated randomized benchmarking, with fidelity as low as $10^{-4}$, commonly considered sufficient for fault-tolerant quantum computing \cite{brown2011single, harty2014high, mount2015error}. This leads us to propose a simple hybrid algorithm where we succeed to reduce the number of qubits needed for the computations by working directly in the Fourier space where the most problematic part --- the shift operator --- is diagonal (see Methods section for more details). In this space, each spin component of the quantum walks evolves independently from the others, allowing to separate the computations on the maximum number of fault-tolerant qubits available. In the case of a perfect quantum computer, one can directly used the full-quantum scheme proposed by A. Shakeel \cite{shakeel2020efficient}. In the other case, one obtains a hybrid scheme where transformation to Fourier space is performed classically (via the Fast Fourier Transform(FFT)) at the beginning of the algorithm, and its inverse is performed, also classically, at the end of the algorithm. In between, the quantum operations are performed on the different set of fault-tolerant qubits before measuring the final wave-function. The minimum number of necessary qubits to perform this hybrid scheme is two, allowing computations to be executed on current NISQ devices. In the following, we choose to develop the numerical scheme in this limit where the errors are small enough to get meaningful results.

\begin{figure}[ht]
\centering
\includegraphics[height=0.3\linewidth]{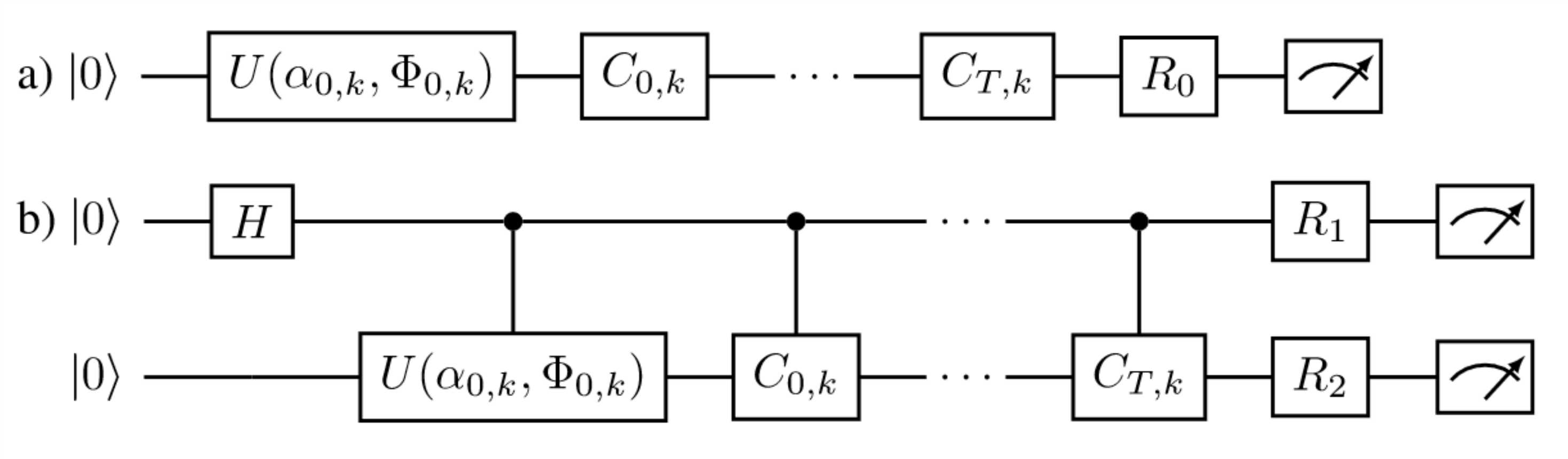}
\caption{Quantum part of the hybrid algorithm for DTQW to compute every $\ket{k}_T$.}
\label{fig:Quantum Circuit}
\end{figure}

The algorithm is composed of two distinct parts. In the first one we perform the FFT on the initial classical state $\psi_{0,p}$, where $p$ refers to the discrete space $p\in \mathbf{N}_p=\{- N/2,- N/2+1,..., N/2-1\}$ and $N$ is a power of $2$, in order to get the Fourier-transformed $\hat{\psi}_{0,k}$ with $k\in \mathbf{N}_p$. In Fourier space, each $\hat{\psi}_{l,k}$, with $l \in \mathbf{N}$ the discrete-time coordinate, evolves independently from the others, allowing us to parallelise the computation of each mode. Moreover, $\forall k,$ we need to memorize the normalization factor $n(k)=\sqrt{|\hat{\psi}_{0,k}^L|^2+|\hat{\psi}_{0,k}^R|^2}$ and the global phase $\Phi_k^+$ for further steps of the algorithm. In order to apply the quantum circuit, we need to encode the above classical information in a quantum state; this can be done efficiently, as follows. At the beginning, for each mode, the quantum state, represented by a qubit, is set to $\ket{0}$ in the canonical basis. Then, we perform a quantum rotation in the Bloch sphere: 
\begin{eqnarray}
\ket{k}_0 = U(\alpha_{0,k},\Phi_{0,k}^-)\ket{0},
\end{eqnarray}
where
\[
U(\alpha_{0,k},\Phi_{0,k}^-)=\begin{pmatrix}
    \cos(\alpha_{0,k}/2) &-\sin(\alpha_{0,k}/2)  \\
    \sin(\alpha_{0,k}/2)e^{i\Phi_{0,k}^-} &\cos(\alpha_{0,k}/2)e^{i\Phi_{0,k}^-}
    \end{pmatrix}. 
    \]
    
The encoded initial quantum state finally reads: \begin{equation}
\hat{\psi}_{0,k}=n(k)e^{i\Phi_{0,k}^+}\ket{k}_0.
\end{equation}

As we show in Fig. \ref{fig:Quantum Circuit}, the total evolution of the walker is achieved by the quantum sub-routine \textit{a)}, by performing one quantum rotation $C_{l,k}$ on each $\ket{k}_0$ (see Methods section). After $T$ such rotations, the final qubit reads 
\begin{equation}
\ket{k}_T =e^{i\Phi_{T,k}^+}\begin{pmatrix} \cos(\alpha_{T,k}/2) \\
    \sin(\alpha_{T,k}/2)e^{i\Phi_{T,k}^-} 
    \end{pmatrix}.
\end{equation}

Finally the state is successively measured into the $x$,$y$,$z$-basis by choosing $R_0=H,S^{\dagger}_1H,I_2$, with $H=\frac{1}{\sqrt{2}}\begin{pmatrix}
    1 &1  \\
    1 &-1
    \end{pmatrix}$ and $S_1=\begin{pmatrix}
    1 &0  \\
    0 &i
    \end{pmatrix}$ and by repeating the procedure until one gets enough statistics to determine the coefficients $\alpha_{T,k}$ and $\Phi_{T,k}^-$. However, in order to implement the very last step of the algorithm, namely the inverse FFT, one needs also the global phase $\Phi_{T,k}^+$. This can be done using the quantum circuit \textit{b)}, where a single qubit controls the $C_{l,k}$ rotations, allowing at the end to measure the global phase by choosing $R_1=H,S_1^{\dagger}H$ and $R_2=I_2$.  Finally, we can perform the classical inverse FFT on
\begin{equation}
\hat{\psi}_{T,k}=n(k)e^{i(\Phi_{0,k}^++\Phi_{T,k}^+)}\begin{pmatrix}
    \cos(\alpha_{T,k}/2)  \\
    \sin(\alpha_{T,k}/2)e^{i\Phi_{T,k}^-}
    \end{pmatrix}
\end{equation}
 to obtain the final state of the quantum walk $\psi_{T,p}$.

\subsection*{Hybrid simulations on IBM's quantum processors}

\begin{figure}
     \centering
     \begin{subfigure}[b]{0.23\textwidth}
         \centering
         \includegraphics[width=\textwidth]{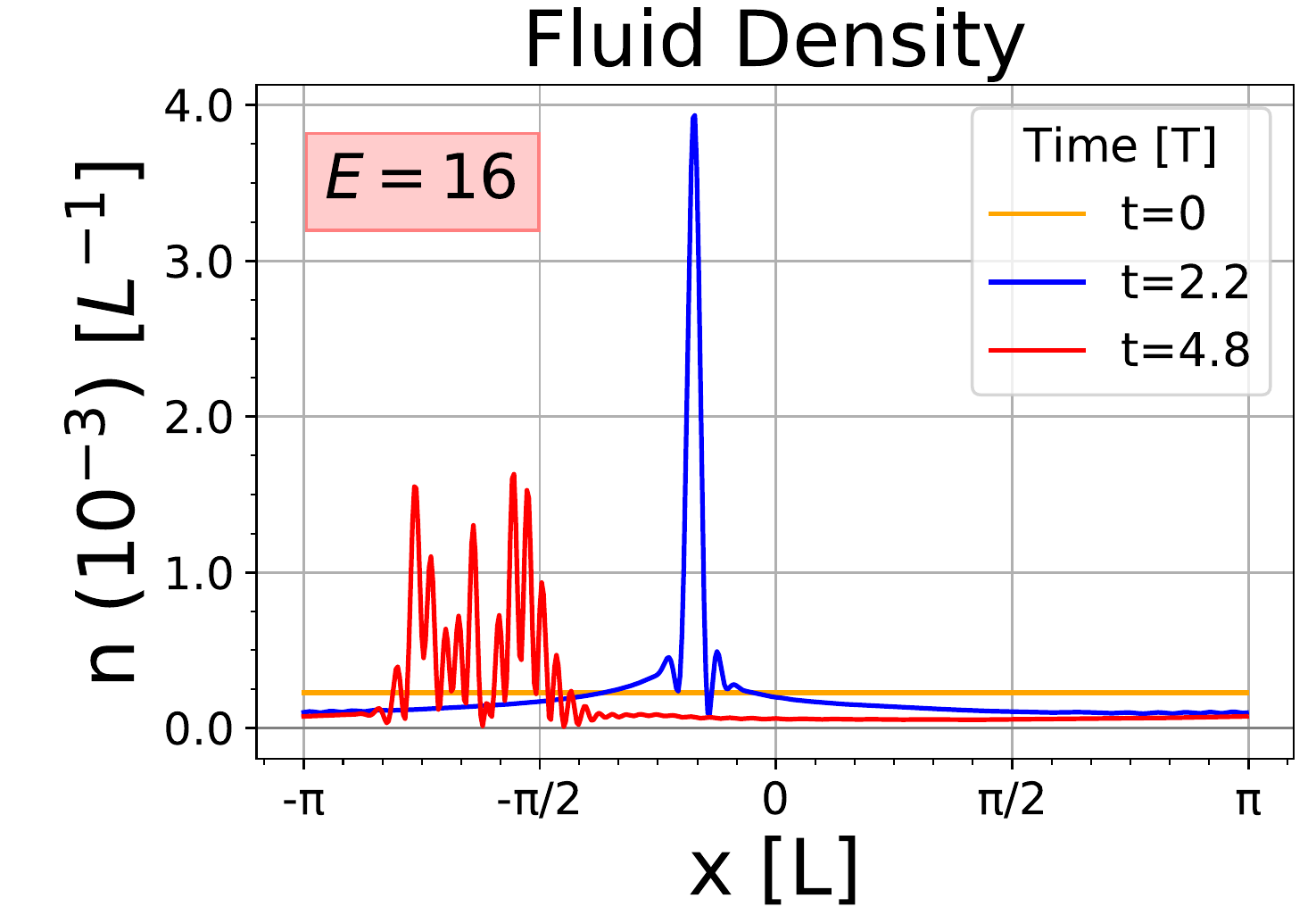}

     \end{subfigure}
     \hfill
     \begin{subfigure}[b]{0.23\textwidth}
         \centering
         \includegraphics[width=\textwidth]{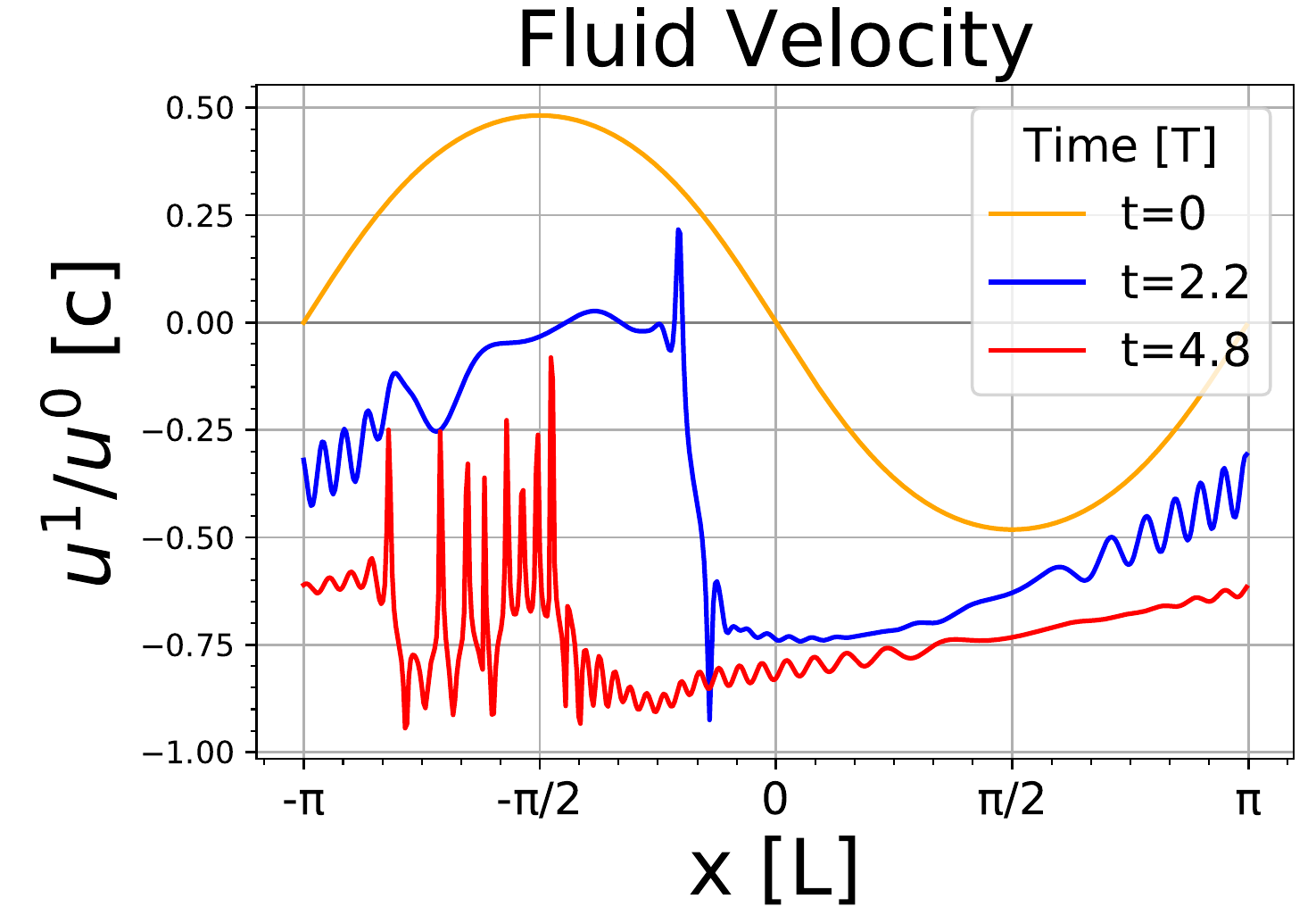}

     \end{subfigure}
    \caption{Profiles of density $n$ (left) and velocity $u^1/u^0$ (right) at different times, for an electric field $E=16$, charge $q=-1$, mass $m=64$, and initial maximum velocity $u_{\text{max}}=0.55$. The mesh size is $N=4096$ and $\epsilon=2\pi/N$.}
\label{profil p for three j}
\end{figure}

\begin{figure}
     \centering
     \begin{subfigure}[b]{0.23\textwidth}
         \centering
         \includegraphics[width=\textwidth]{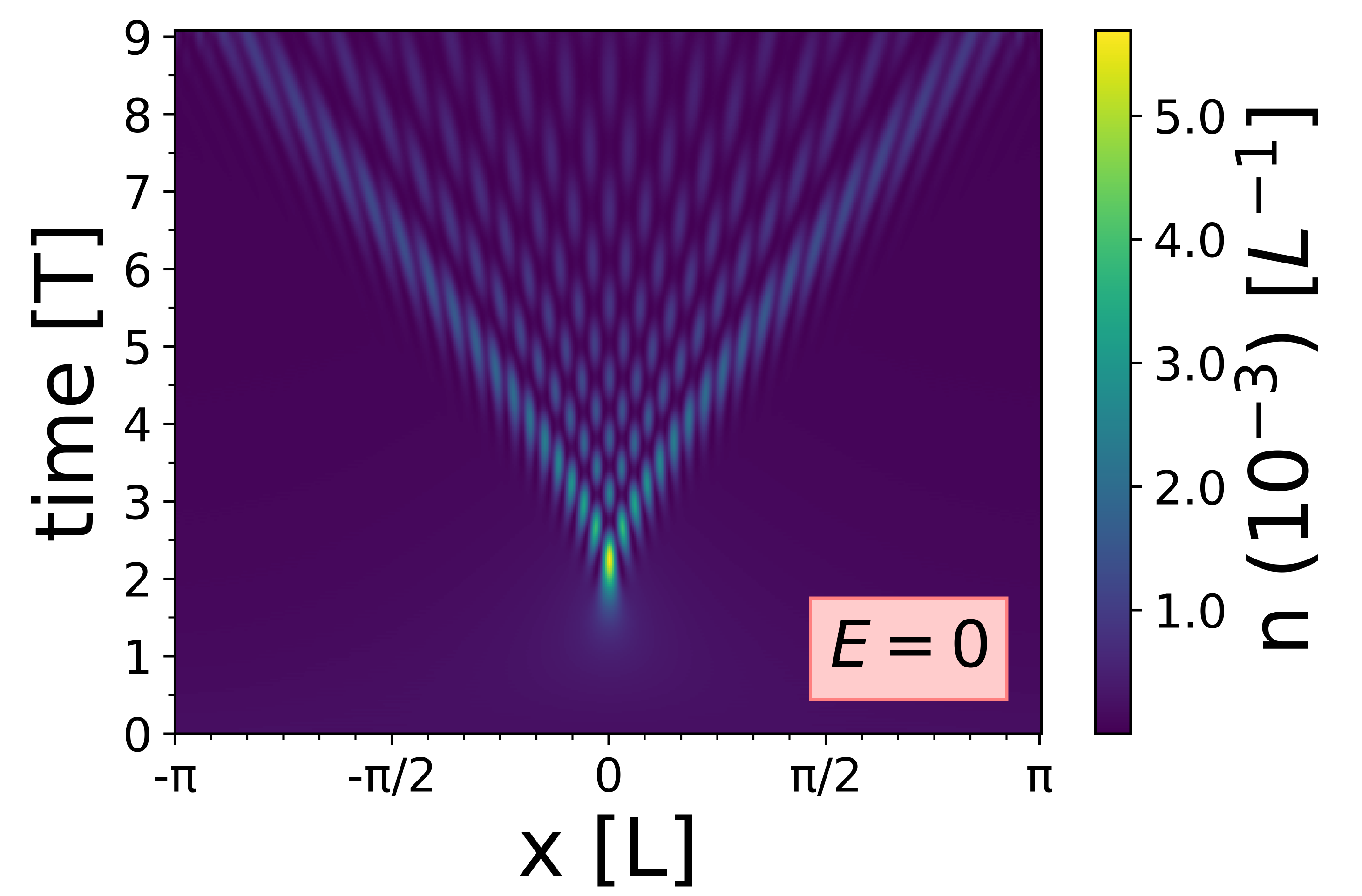}
       
     \end{subfigure}
     \hfill
     \begin{subfigure}[b]{0.23\textwidth}
         \centering
         \includegraphics[width=\textwidth]{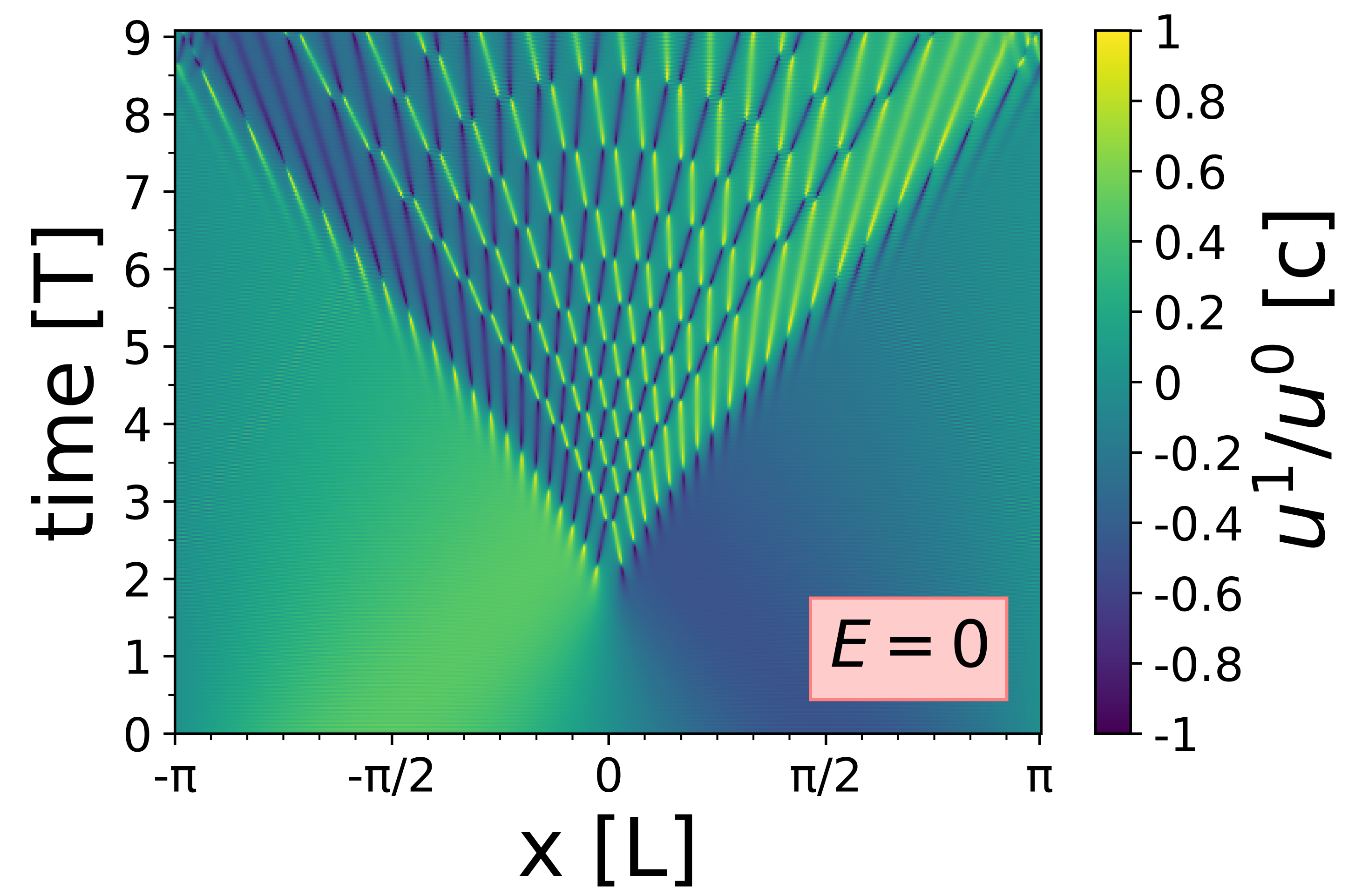}

     \end{subfigure}
     \hfill
     \begin{subfigure}[b]{0.23\textwidth}
         \centering
         \includegraphics[width=\textwidth]{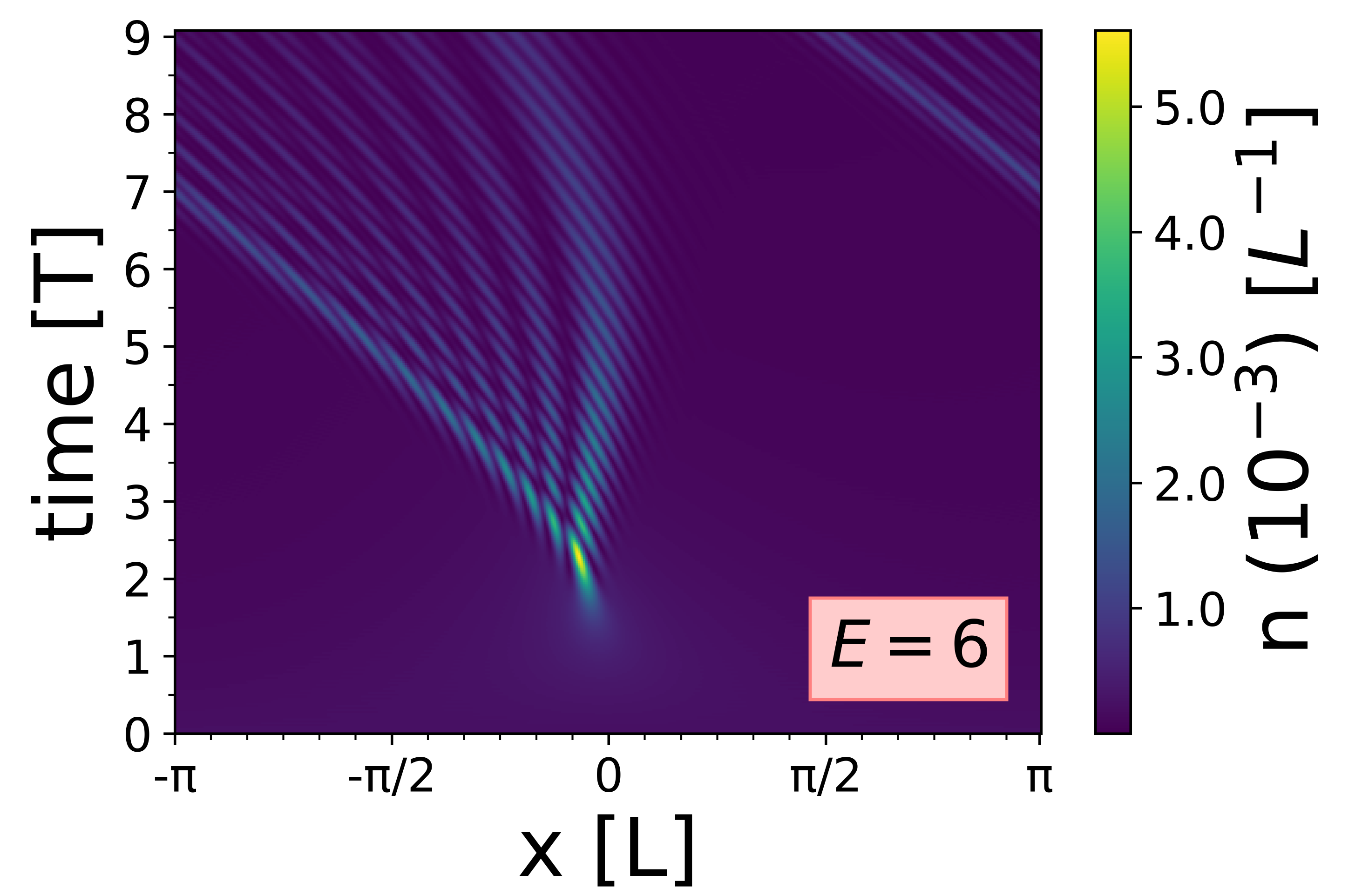}

     \end{subfigure}
     \hfill
     \begin{subfigure}[b]{0.23\textwidth}
        \centering
        \includegraphics[width=\textwidth]{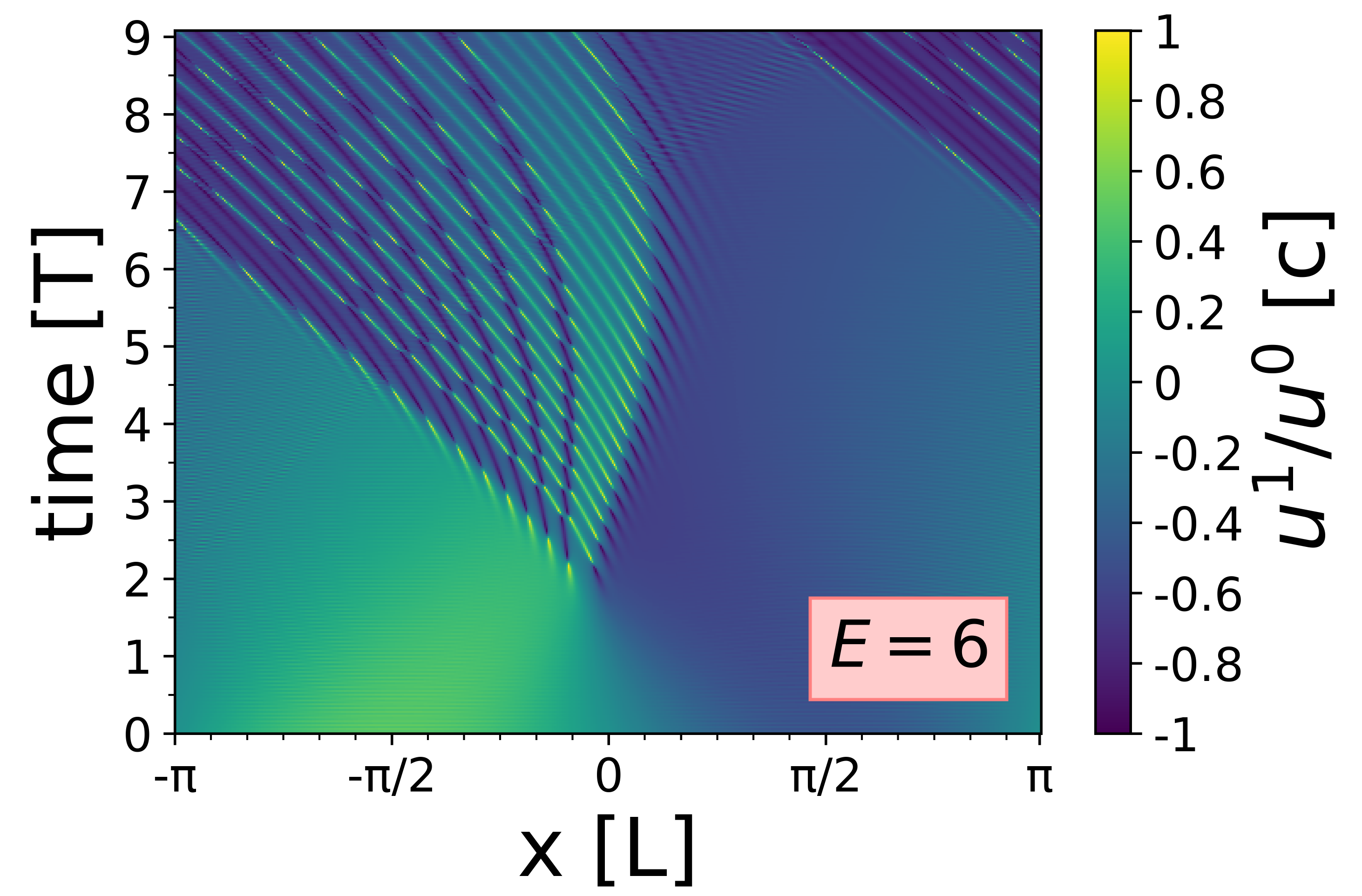}

     \end{subfigure}
     \hfill
     \begin{subfigure}[b]{0.23\textwidth}
         \centering
         \includegraphics[width=\textwidth]{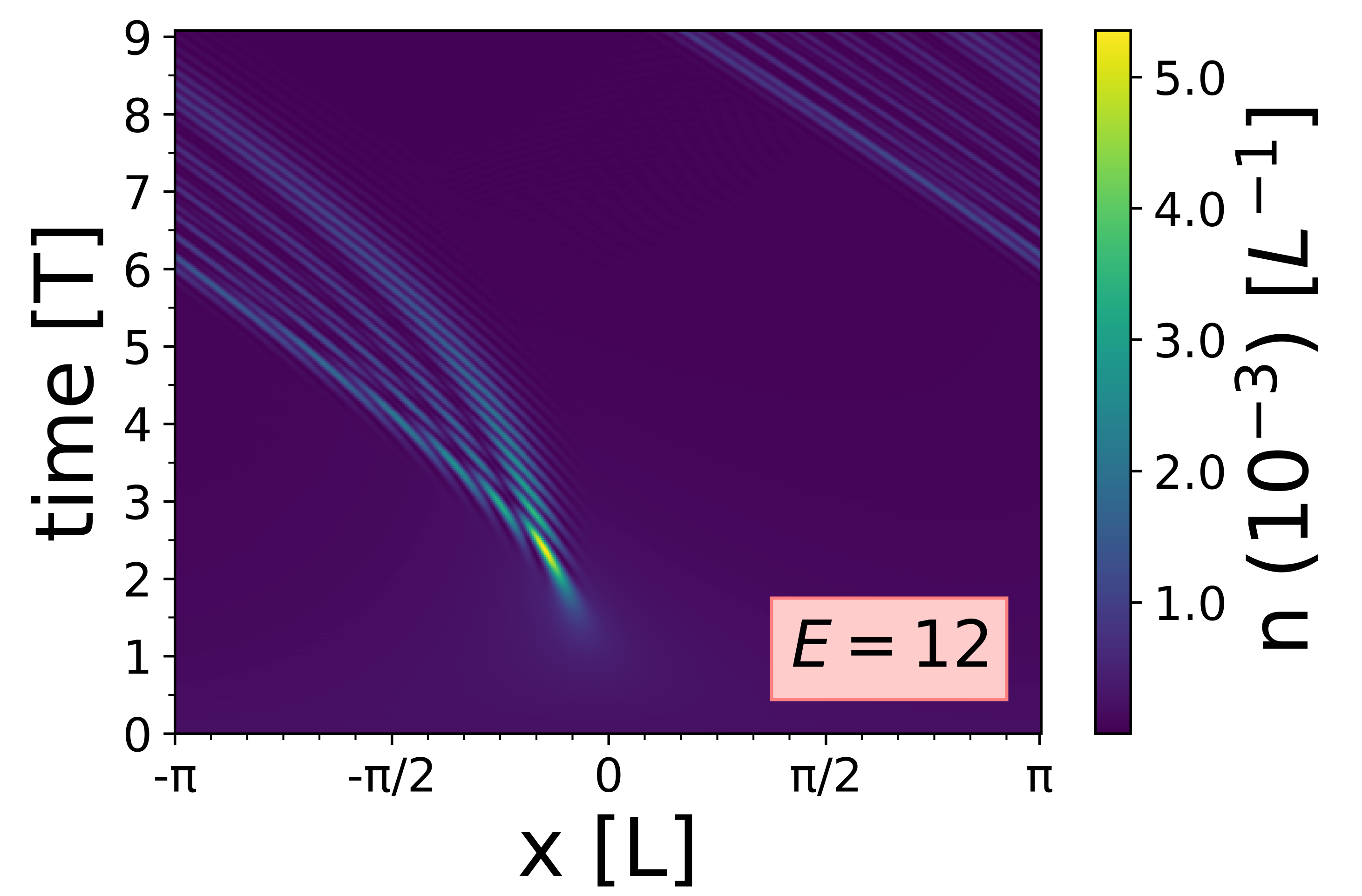}

     \end{subfigure}
     \hfill
     \begin{subfigure}[b]{0.23\textwidth}
        \centering
        \includegraphics[width=\textwidth]{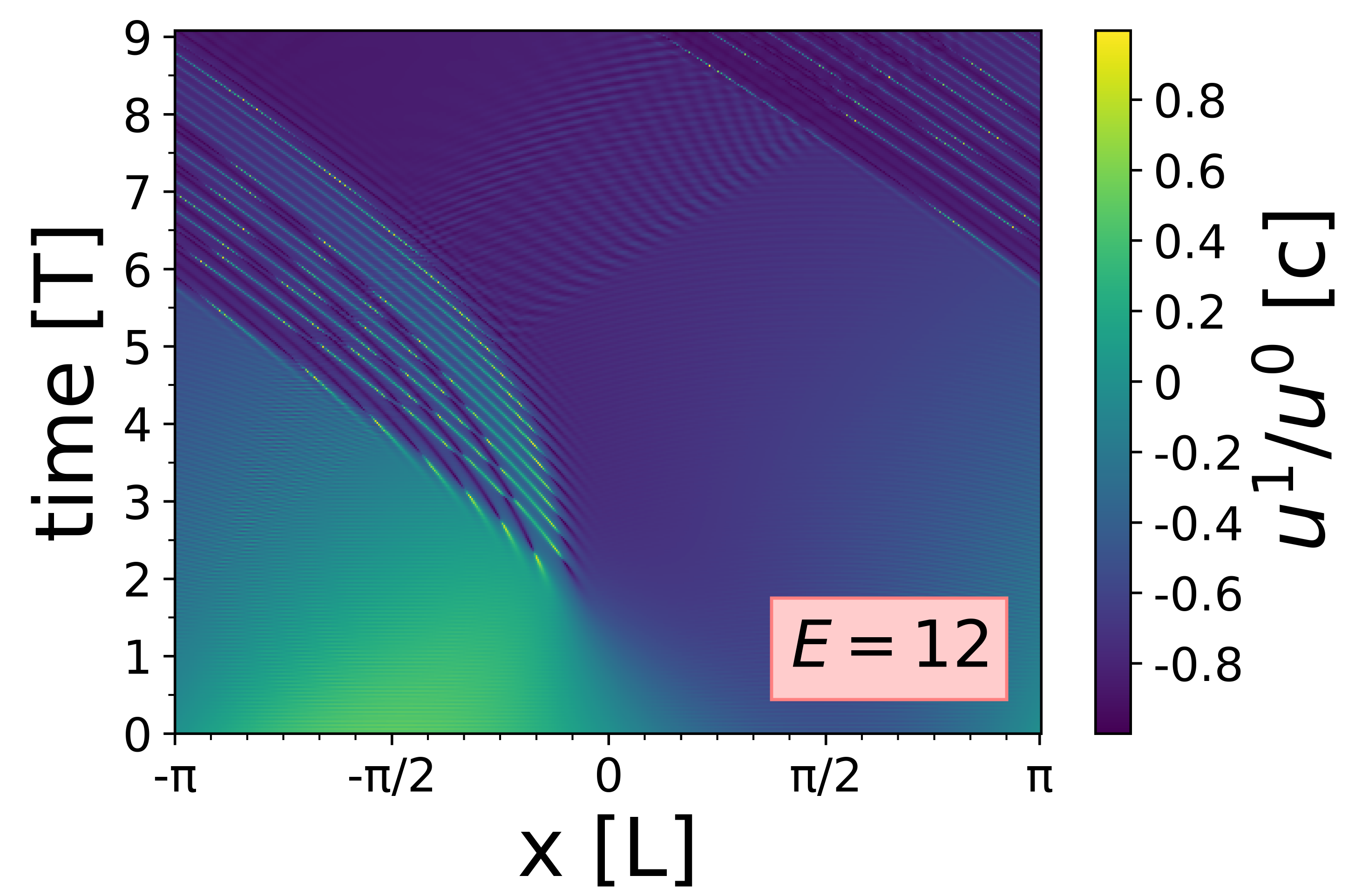}

     \end{subfigure}
    \caption{Evolution of the shock's density $n$ (left) and velocity $u^1/u^0$ (right) for different values of the electric field $E=0,8,12$, charge $q=-1$,  mass $m=64$, and initial maximum velocity $u_{\text{max}}=0.55$. The mesh size is $N=4096$ and $\epsilon=2\pi/N$.}
\label{profils jp}
\end{figure}

\begin{figure}
     \centering
     \begin{subfigure}[b]{0.23\textwidth}
         \centering
         \includegraphics[width=\textwidth]{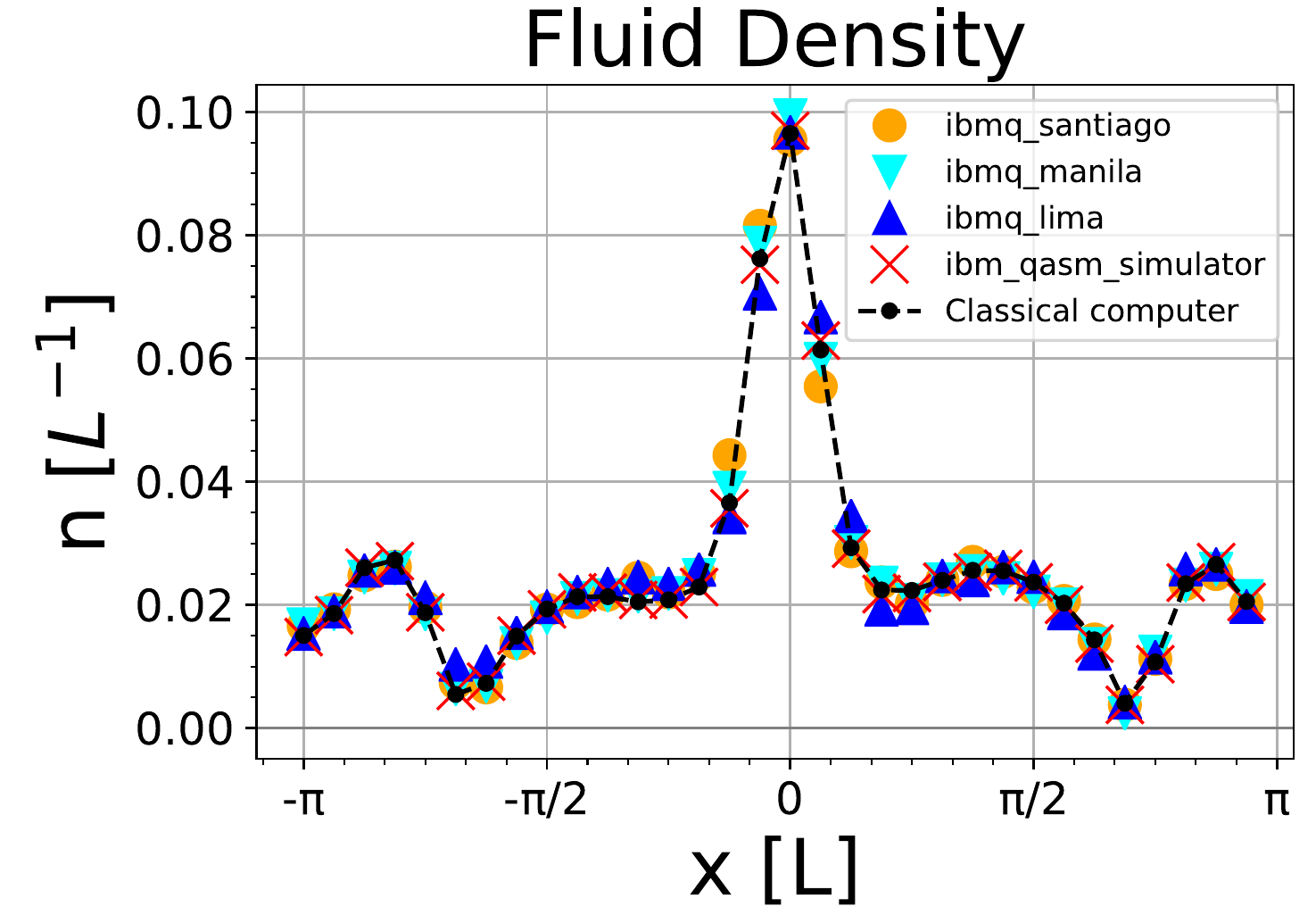}

     \end{subfigure}
     \hfill
     \begin{subfigure}[b]{0.23\textwidth}
         \centering
         \includegraphics[width=\textwidth]{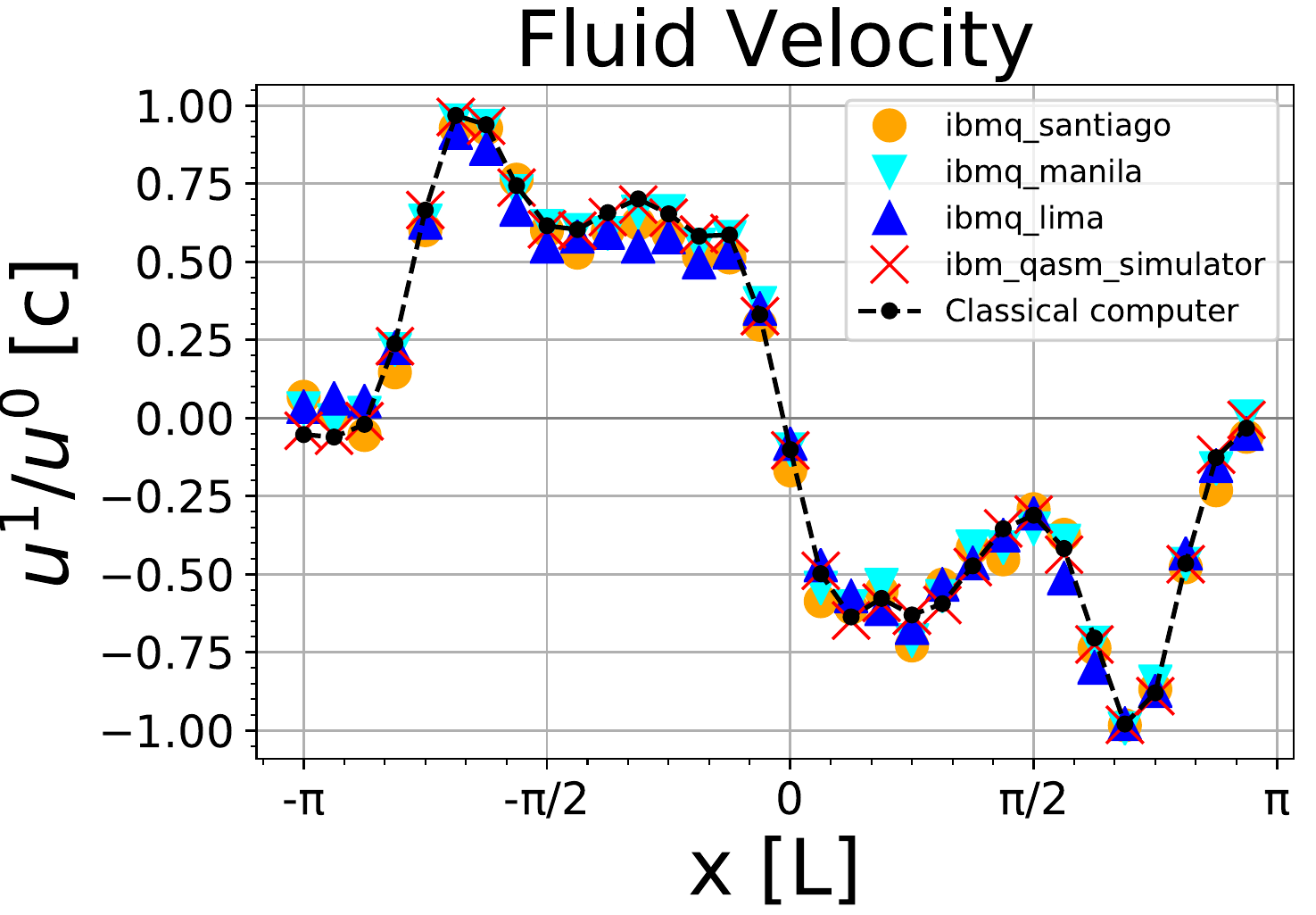}
         
     \end{subfigure}
     \hfill
     \begin{subfigure}[b]{0.23\textwidth}
         \centering
         \includegraphics[width=\textwidth]{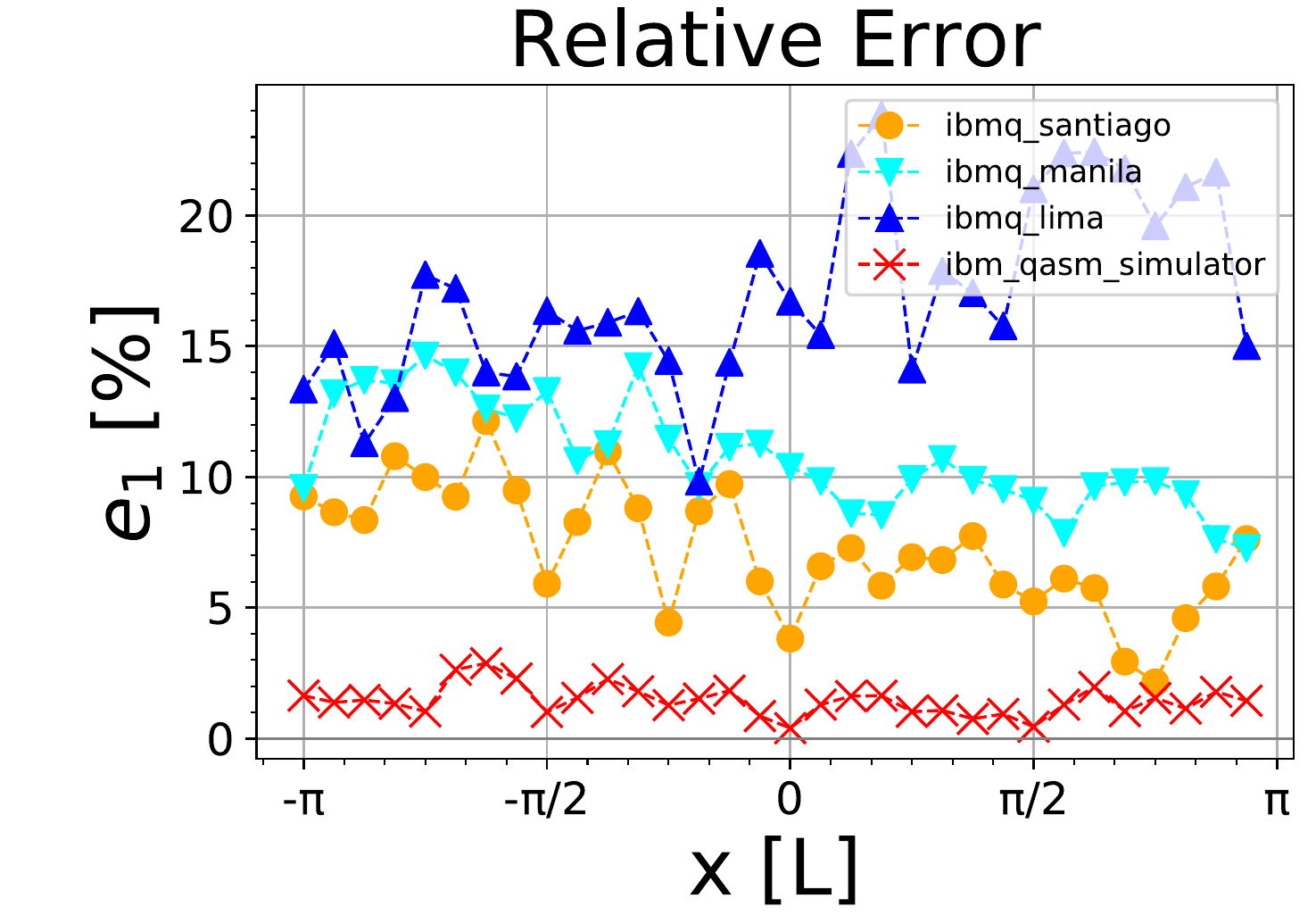}
         
     \end{subfigure}
     \hfill
     \begin{subfigure}[b]{0.23\textwidth}
         \centering
         \includegraphics[width=\textwidth]{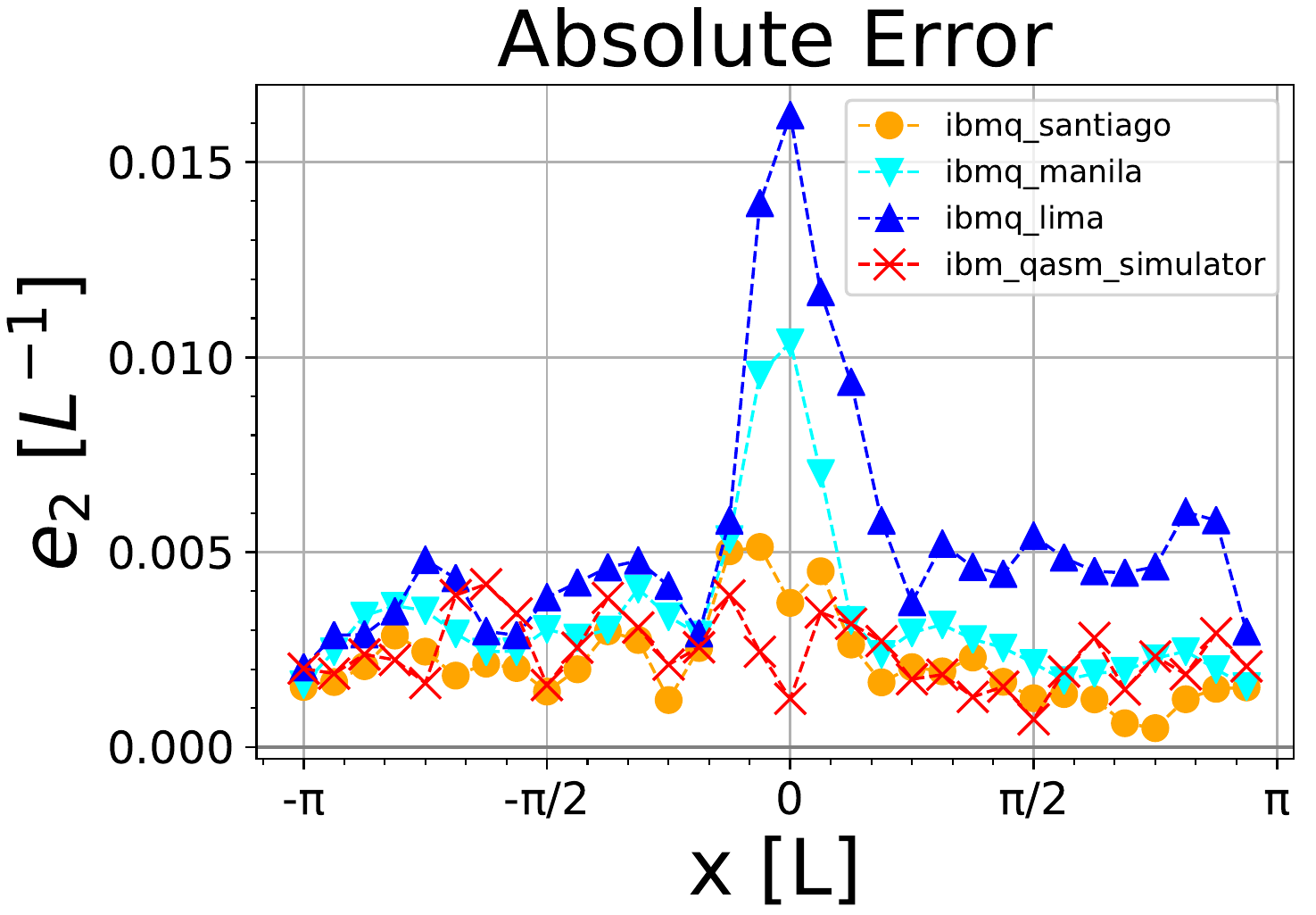}
         
     \end{subfigure}
    \caption{Shock's profiles of fluid density $n$ (upper left), fluid velocity $u^1/u^0$ (upper right) computed on different IBM's quantum processors, {\tt ibm$\_$qasm$\_$simulator} and a classical computer, at $t=1.96$. The lower panels show relative errors (lower left)  and absolute error (lower right)  between the ideal results obtained on the classical computer and the results obtained on the quantum processors and simulator. The simulation parameters are: electric field $E=0.6$, charge $q=-1$, mass $m=6$, and initial maximum velocity $u_{\text{max}}=0.92$. The mesh size is $N=32$, and $\epsilon=2\pi/N$.}
\label{multi ibmq}
\end{figure}

\begin{figure}
     \centering
     \begin{subfigure}[b]{0.23\textwidth}
         \centering
         \includegraphics[width=\textwidth]{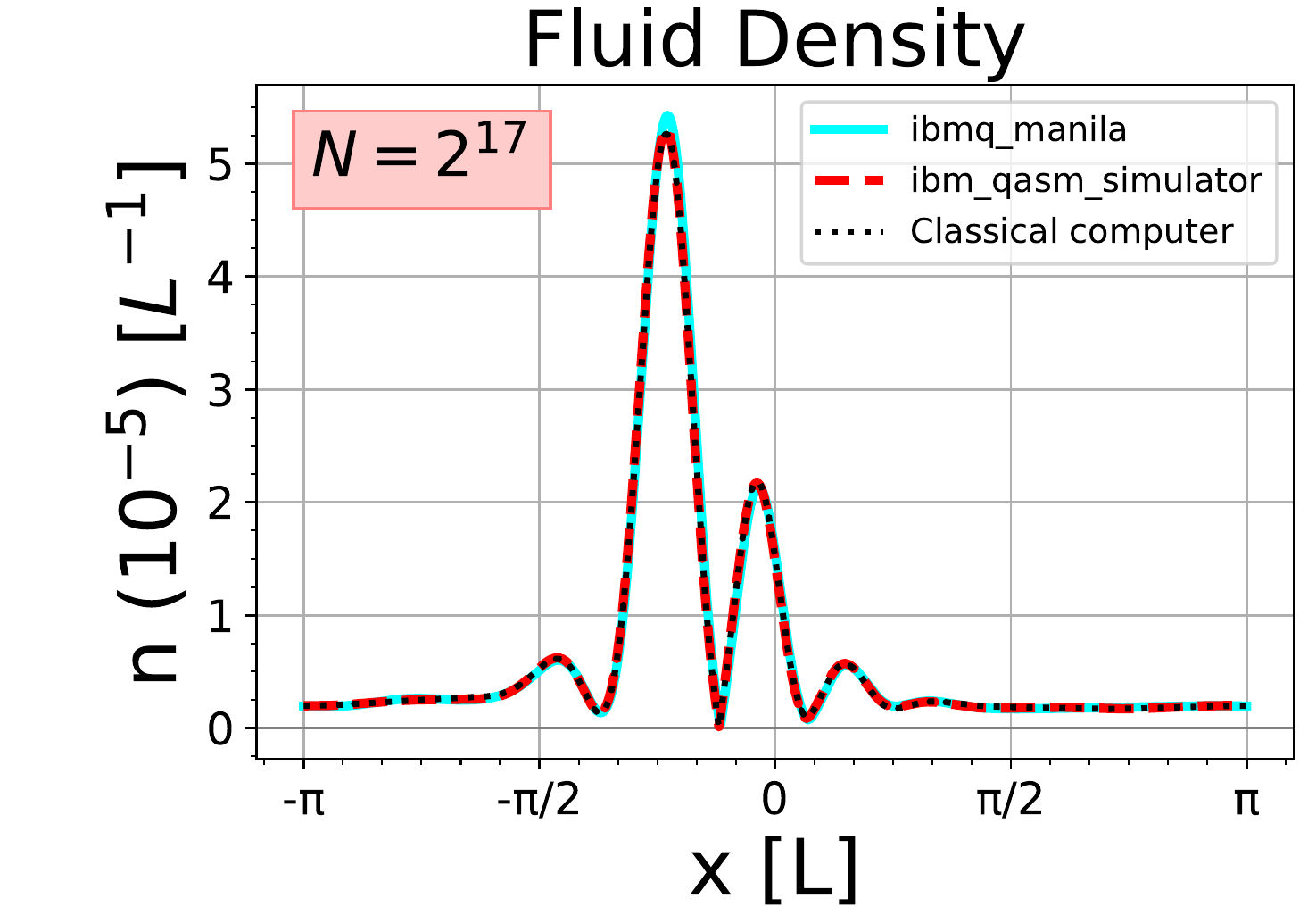}

     \end{subfigure}
     \hfill
     \begin{subfigure}[b]{0.23\textwidth}
         \centering
         \includegraphics[width=\textwidth]{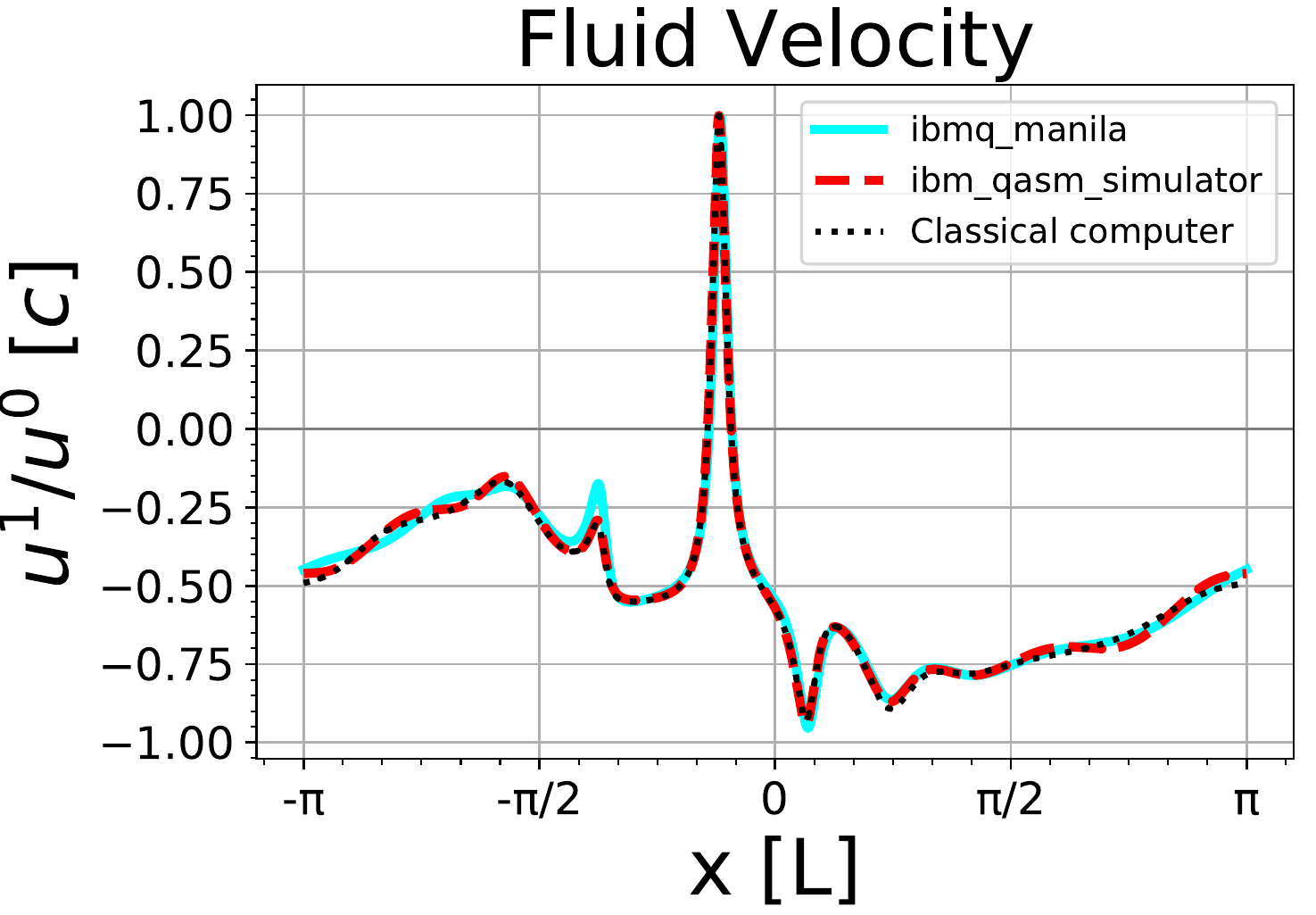}
         
     \end{subfigure}
     \hfill
     \begin{subfigure}[b]{0.23\textwidth}
         \centering
         \includegraphics[width=\textwidth]{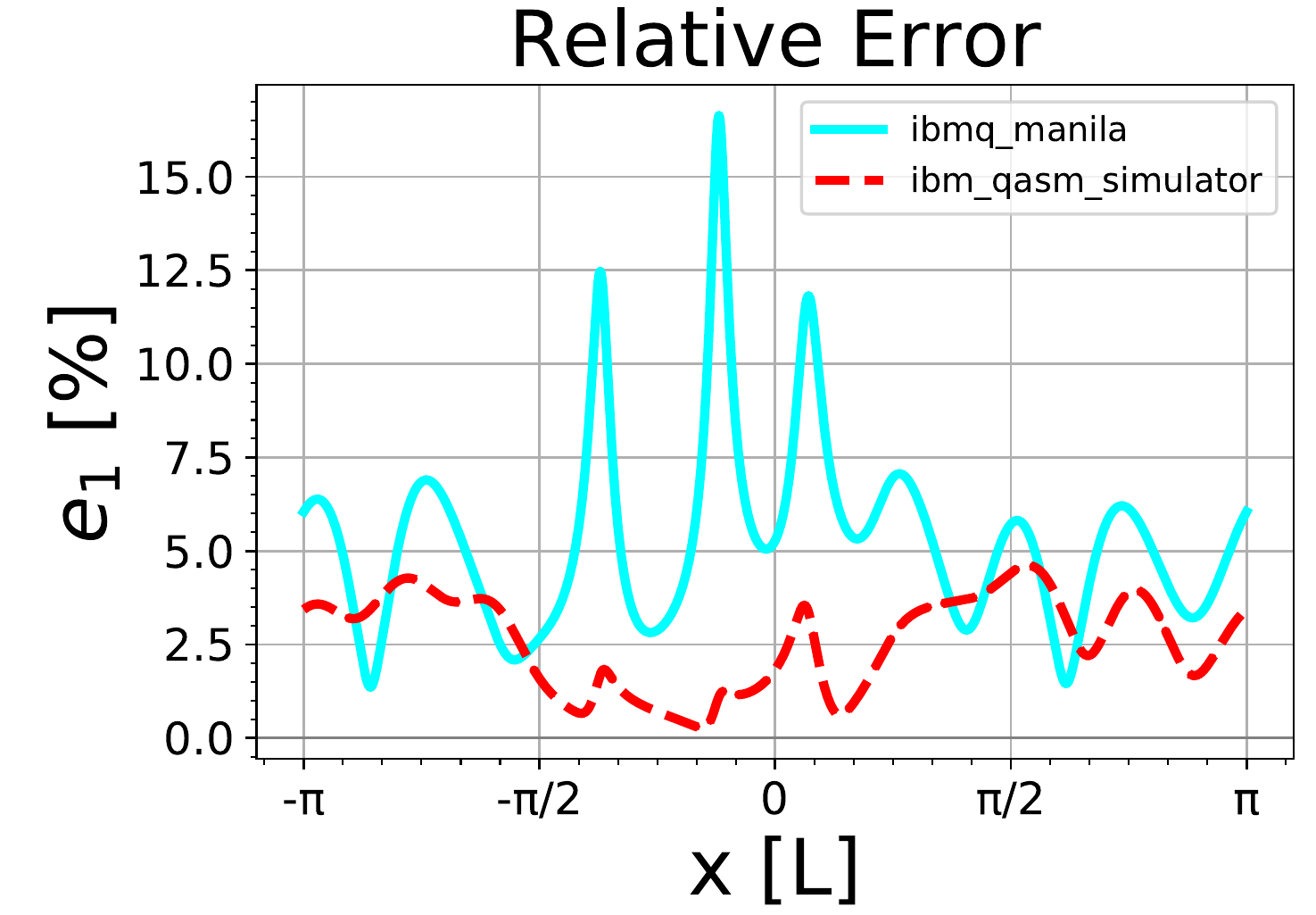}
         
     \end{subfigure}
     \hfill
     \begin{subfigure}[b]{0.23\textwidth}
         \centering
         \includegraphics[width=\textwidth]{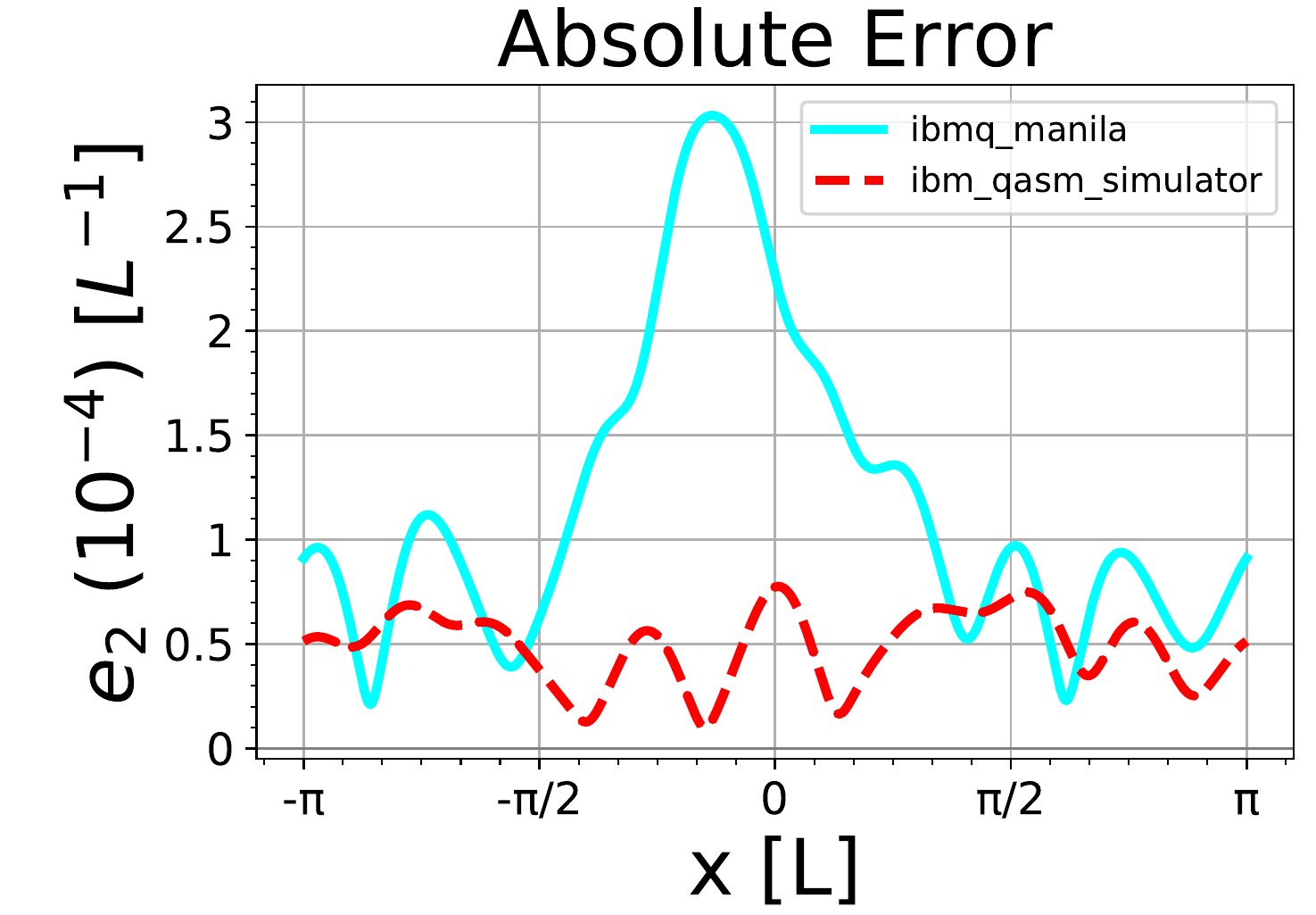}
         
     \end{subfigure}
    \caption{High resolution shock's profiles of fluid density $n$ (upper left), fluid velocity $u^1/u^0$ (upper right) computed on {\tt ibmq$\_$manila} quantum processor, {\tt ibm$\_$qasm$\_$simulator} and classical computer at $t=2.5$, with $N=2^{17}$ grid points, an electric field $E=2$, a charge $q=-1$, a mass $m=6$, an initial maximum velocity $u_{\text{max}}=0.92$, $\epsilon=2\pi/N$ and relative errors (lower left)  and absolute error (lower right)  between the ideal results obtained on a classical computer and the results obtained on quantum processors and simulator.}
\label{IBM q N=2^17}
\end{figure}

Simulations have been performed on classical processors (Figures \ref{profil p for three j} and \ref{profils jp}), and on IBM's publicly available quantum processors (Figures \ref{multi ibmq} and \ref{IBM q N=2^17}). The initial condition of the quantum walk is chosen such as to obtain hydrodynamical shocks: the initial fluid density $n$ is constant, while the initial fluid velocity $u^1/u^0$ is anti-symmetric with respect to $x=0$ (see Methods for more details). 
In Figure \ref{profil p for three j}, the fluid density and velocity are displayed at three different times. The shock is characterised by a peak in the fluid density $n$ and a small region with a large gradient in the fluid velocity $u^1/u^0$ at t=2.2. After the impact, the front of the shock propagates to the left due to the external electric field, yielding a non-trivial shock structure at t=4.8. Figure \ref{profils jp} shows the fluid density and velocity with respect to space and time for several values of the electric field. The shocks are perfectly symmetric around $x = 0$ in the absence of electric field. For non vanishing electric fields, the shocks are accelerated in the direction of the field. 
These results have been successfully recovered using IBM's quantum processors. Figure \ref{multi ibmq} shows the first simulations of hydrodynamical shocks using NISQ devices on a line of $N=32$ nodes. The same simulation has been performed on three different quantum processors ({\tt ibmq$\_$santiago}, {\tt ibmq$\_$manila}, {\tt ibmq$\_$lima}) and the results are compared with a simulator of quantum devices ({\tt ibm$\_$qasm$\_$simulator}) and a classical computer. The performances of the different IBM's quantum processors are compared with the relative error defined as
\begin{equation}
e_1=100\frac{\sqrt{|\psi^L_{x,q}-\psi^L_{x,c}|^2+|\psi^R_{x,q}-\psi^R_{x,c}|^2}}{\sqrt{|\psi^L_{x,c}|^2+|\psi^R_{x,c}|^2}},
\end{equation} and the absolute error defined as
\begin{equation}e_2=\sqrt{|\psi^L_{x,q}-\psi^L_{x,c}|^2+|\psi^R_{x,q}-\psi^R_{x,c}|^2},
\end{equation}
where $q$ refers to the quantum devices and simulator and $c$ to the classical computer. Even if the relative errors range from $3\%$ to $24\%$, the results on the fluid density and velocity are qualitatively accurate, showing the expected shock. The finite number of measurements $M=8096$ leads to statistical errors of the order of $3\%$ as shown by the relative errors of {\tt ibm$\_$qasm$\_$simulator}.
Figure \ref{IBM q N=2^17} shows the results obtained on a grid of $N=2^{17}$ points with the {\tt ibmq$\_$manila} quantum processor showing first that this hybrid algorithm allows to perform large simulations on NISQ devices. The velocity almost reaches the speed of light ${u^1}/{u^0}\approx 0.9993$ at $x\approx -\frac{3\pi}{24}$ where the density $n$ nearly vanishes, demonstrating ultra-relativistic effects in the shocks.

\section*{Discussion}

Let us now discuss the results presented in this article,
focusing in particular on possible extensions. 

All the results presented above address hydrodynamics in $(1 + 1)$ space-time dimensions, and should therefore be extended to higher dimensions. We believe this extension should be possible but is non-trivial either because (i) the number of spinor components depends on the space-time dimension ({\sl e.g.} a spinor in $(1 + 3)$ dimensions), and (ii) the Madelung transformation for a non-charged Dirac fluid is much simpler in $(1 + 1)$ dimensions than in higher dimensions. 

In higher dimensions, a charged fluid can be coupled not only to electric fields, but also to magnetic ones. Moreover, these fields may not  be uniform and constant, as is the electric field considered in this work. 
More generally, Dirac particles and their discrete counterparts, {\sl i.e.}, DTQWs, can also be coupled to arbitrary Yang-Mills gauge field \cite{QWnonabelian,arnault:hal-02318234} and to relativistic gravitational fields  \cite{PhysRevA.88.042301,Arnault:2016uol,condmat4020040}. Extending the above results in these directions would for example open up the possibility of simulating quark-gluon plasmas and extreme astrophysical plasmas on hybrid quantum-classical computers.
DTQWs can also be used as basis to build full fledged discrete gauge theories. Can the hybrid algorithm presented above be extended to simulate these discrete gauge theories? If so, the extension would make it possible to simulate not only fluids in external, imposed gauge fields, but also self-consistent problems where the dynamics of the gauge fields and the fluids are fully coupled, as in self-consistent plasma dynamics.

Finally, the hybrid algorithm we propose should be extended to arbitrary quantum and classical fluids. The key is to consider more general quantum walks and quantum automata. Self-interacting walks and automata  \cite{DMDB15a,Lubasch_2020,PhysRevA.103.052210} are of particular interest because they are a relatively easy tool to model arbitrary equations of state and can probably be realized using Bose-condensates \cite{Bunkov_2020,Kjaergaard_2020}.

\section*{Conclusion}

We have shown that present-day IBM's NISQ devices can simulate quantum-relativistic-charged fluids in an electric field. Our approach is based on a novel hybrid classical-quantum algorithm using DTQW with continuous-limit Dirac equation mapped into relativistic hydrodynamics by a generalization of the Madelung transformation. We have also discussed several extensions which may reasonably be carried out with success in the near future. These include non-quantum fluids and fluids coupled to arbitrary gauge fields. All in all, this work opens the door to more efficient quantum simulation of quantum and classical hydrodynamics \cite{Wyatt,TSUBOTA2013191}, with natural applications to quantum, possibly relativistic plasmas \cite{manfredi2005model,haas2011quantum,Bonitz}.

\section{Methods}

\paragraph{Quantum walks as discretisations of the Dirac equation.}

DTQWs are defined in discrete space and discrete time and have an internal degree of freedom usually called the coin. In this article, we focus on DTQWs defined in discrete $1D$
space. Having spectral simulations in mind, we also take space to be $N$-periodic, where $N$ is a power of 2 and we label the grid points by $p \in \mathbf{N}_p =\{- N/2,- N/2+1,..., N/2-1\}$. Discrete instants are labelled by $l\in \mathbf{N}$. We also choose the coin-space to be $2D$ and denote by $(\ket{L}, \ket{R})$ an arbitrary fixed orthonormal basis in that space. 
With these conventions, 
the state of the walk at time $l$ can be written as $\ket{\psi}_{l}=\sum_{p}\psi_{l,p}^L\ket{p}\ket{L}+\psi_{l,p}^R\ket{p}\ket{R}$
where the set of complex numbers $\{\psi^L_{l,p}, \psi^R_{l,p}\}$ with $p \in \mathbf{N}_p$ represents the two-component wave-function of the walk at time $l$.
At each time-step, the walk is advanced through the successive action of two unitary operators, one which acts in position-space and one which acts in coin-space. The operator 
${\hat S}$ acts in position-space and is usually called the shift operator; it is defined by $\hat{S}=\ket{L}\bra{L}\sum_{p}\ket{p-1}\bra{p}+\ket{R}\bra{R}\sum_{p}\ket{p+1}\bra{p}$. The shift operator is thus a coin-conditioned spatial translation which moves every $\psi_{l,p}^L$ to the left by one unit and every $\psi_{l,p}^R$ to the right, also by one unit. 
The operator ${\hat C}_l$ acting in coin-space is allowed to depend on time $l$ and, at each point $p$, mixes the $L$ and $R$ components in a unitary manner. 
This operator is defined by ${\hat C_l}=\sum_{p}\hat{C}_ {l,p}\ket{p}\bra{p}$ with $\hat{C}_{l,p}=e^{-i\epsilon q(A_0)_{l,p}}R_X(2\epsilon m)R_Z(-2\epsilon q(A_1)_{l,p})$, where $R_X(\theta)=\begin{pmatrix} 
    \cos{(\theta/2)} & -i\sin{(\theta/2)}  \\
    -i\sin{(\theta/2)} & \cos{(\theta/2)}
    \end{pmatrix}$ and $R_Z(\theta)=\begin{pmatrix} 
    e^{-i\theta/2} & 0  \\ 0 & e^{i\theta/2}
\end{pmatrix}$ are primitive single qubit operations \cite{nielsen2002quantum}. 
The potential vector $A_0$ and $A_1$ are arbitrary real numbers, as are the two real positive parameters $\epsilon$ and $m$. It is useful to introduce the notation $\hat{U}_l=\hat{C}_l\hat{S}$, which makes it possible to write the evolution equation of the quantum walks in the compact form $\ket{\psi}_{l+1}=\hat{U}_l\ket{\psi}_{l}$. The interpretation of these quantities becomes clear in the continuum limit.
The continuum limit can be investigated by introducing the space-time coordinates $x_p=\epsilon p$, $t_l=\epsilon l$ and letting $\epsilon$ tend to zero \cite{DIMOLFETTA2014157}. The wave-function of the walk then becomes a continuous function of $x$ and $t$ which obeys the Dirac equation introduced earlier. 

\paragraph{Spectral formulation.} To make the computation simpler, we choose the gauge $A_0=0$, $(A_1)_{l}=El\epsilon$ where the vector potential depends only on the discrete time $l$ and so the coin operator $\hat{C}_{l,p}=\hat{C}_l$.
The classical and hybrid simulations are accomplished in Fourier space where the shift operator entering the definition of the walks amounts to a coin-controlled
multiplication by a phase factor. More precisely, let
$\hat{\psi}_{l,k}=\frac{1}{\sqrt{N}}\sum_{p=- N/2}^{ N/2-1}\psi_{l,p}e^{-2i\pi kp/N}$ be the discrete Fourier transform of a function defined on the
discrete space-time grid. In Fourier space, the equations of the walk, $\forall l,k \in \mathbf{N} \times \mathbf{N}_p$, read:

\begin{eqnarray}
   \begin{pmatrix}
    \hat{\psi}_{l+1,k}^L  \\
    \hat{\psi}_{l+1,k}^R
    \end{pmatrix}
    =C_l
    \begin{pmatrix}
    e^{2i\pi k/N} & 0  \\
    0 & e^{-2i\pi k/N}
    \end{pmatrix}
     \begin{pmatrix}
    \hat{\psi}_{l,k}^L  \\
    \hat{\psi}_{l,k}^R
    \end{pmatrix},
\label{discret eq fourier space}
\end{eqnarray}
where the absence of spatial convolution is due to the choice of gauge.

\paragraph{Initial condition for the simulation of shocks.}
Let us note $\psi(x,t)=e^{i\phi_+/2}\begin{pmatrix} |\psi^L|e^{i\phi_-/2} \\ |\psi^R|e^{-i\phi_-/2} \end{pmatrix}$ with  $|\psi^L|=\frac{1}{\sqrt{2}}\sqrt{j^0-j^1}$ and $|\psi^R|=\frac{1}{\sqrt{2}}\sqrt{j^0+j^1}$. In order to get a shock we need an anti-symmetric initial velocity $u^1/u^0$, thus we choose a global phase $\phi_+=2mu_{\text{max}}\cos(x)$ with ${\text{max}}$ a positive number and a relative phase $\phi_-=0$. Equation (\ref{abs2}) implies that $j^1=-nu_{\text{max}}\sin(x)$ and so $u^1=-u_{\text{max}}\sin(x)$. Then $j^0=+\sqrt{n^2+(j^1)^2}=n\sqrt{1+(u_{\text{max}}\sin(x))^2}$ and finally $u^1/u^0=j^1/j^0$ is anti-symmetric. The fluid density can be an arbitrarily chosen constant; we set $n=1$. These initial condition is inspired by similar choices used to simulate the dynamics of a non-quantum cosmological fluid \cite{Coles2003}, Bose-Einstein condensates of axions \cite{Sikivie2009} and quantum walk hydrodynamics \cite{Hatifi2019}.

\paragraph{Discrete fluid density and velocity.}
Following the previous definition, the current j can be determined from the wavefunction of the DTQW using the formula $(j^0)_{l,p} = |\psi_{l,p}^R|^2+|\psi_{l,p}^L|^2$ and $(j^1)_{l,p} = |\psi_{l,p}^R|^2-|\psi_{l,p}^L|^2$ where $l$ denotes a discrete time coordinate and p a discrete space coordinate. Thus the fluid density reads $n_{l,p}=\sqrt{(j^0)_{l,p}^2-(j^1)_{l,p}^2}=2|\psi_{l,p}^L||\psi_{l,p}^R|$ and the fluid velocity, in $c$ units, $(\frac{u^1}{u_0})_{l,p}=\frac{(j^1)_{l,p}}{(j^0)_{l,p}}$.

\paragraph{Practical implementation on IBM's NISQ devices.} 
Before any computation on its quantum processors, IBM automatically transpiles the quantum circuit in order to reduce the number of primitive quantum operations and the errors. However, the transpiler does not perform efficiently in the case of circuit b) presented Figure \ref{fig:Quantum Circuit}, giving completely noisy results. We found that this difficulty can be overcome if we transpile the quantum circuit a) before transpiling the control-circuit a) which is contained in the circuit b).
\paragraph{Compression of the wavefunction in Fourier space.} 
Figure \ref{IBM q N=2^17} shows a simulation on a grid of $N=2^{17}$ points. This simulation has been successfully performed thanks to a compression of the wavefunction in Fourier space. Indeed, the momentum is bounded by the quantity $mu_{\text{max}}$ and the Fourier space is discretized with a step $\Delta k=\frac{2\pi}{N\Delta x}$. By choosing $\Delta x=\frac{2\pi}{N}$, then $\Delta k=1$ and most of the Fourier components of the DTQW vanishes for $k>k_{\text{max}}=mu_{\text{max}}$ ($\hbar=1$, $c=1$), reducing drastically the computations.

\section*{Acknowledgements}
NFL was partially funded by award no. DE-SC0020264 from the U.S. Department of Energy.
We acknowledge the use of IBM Quantum services for this work. The views expressed are those of the authors, and do not reflect the official policy or position of IBM or the IBM Quantum team.

\section*{Competing interests}
The authors declare no competing interests.

\bibliographystyle{ieeetr}
\bibliography{maintext}
\include{bibliography} 

\newpage

\section*{Supplementary information}

\subsection*{Non-relativistic limit}

\subsubsection{Dirac equation}
The Dirac equation $i\gamma^{\mu}D_{\mu} \psi-m\psi=0$ 
reads, in component terms and in units where $\hbar \ne 1$ $c \ne 1$:

\begin{eqnarray}
\frac{1}{c} (\partial_t+i\frac{qV}{\hbar}) \Psi^{L} - (\partial_x+i\frac{qA_1}{\hbar}) \Psi^{L} & = & -i\frac{mc}{\hbar} \Psi^{R}, \nonumber \\
\frac{1}{c} (\partial_t+i\frac{qV}{\hbar}) \Psi^{R} + (\partial_x+i\frac{qA_1}{\hbar}) \Psi^{R} & = & -i\frac{mc}{\hbar} \Psi^{L}.
\end{eqnarray}

Each component obeys the same Klein-Gordon (KG) equation:
\begin{equation}
\frac{1}{c^2} D_{tt} \Psi^{L/R} - D_{xx} \Psi^{L/R} = - \frac{m^2 c^2}{\hbar^2} \Psi^{L/R},
\end{equation}
where $D_{tt}=(D_t)^2=(\partial_t+i\frac{qV}{\hbar})^2$ and $D_{xx}=(D_x)^2=(\partial_x+i\frac{qA_1}{\hbar})^2$

To determine the non-relativistic limit, we have to extract out the relativistic part of the wavefunction as $\Psi^{L/R} = {\bar \Psi}^{L/R} \exp(-i \frac{mc^2}{\hbar} t)$ and consider ${\bar \Psi}^{L/R}$ as the wavefunction of the particle in the non relativistic limit. The non relativistic energy of the particle is $E'=E-mc^2$ with $E'\ll mc^2$. We therefore expect that $|\frac{\partial {\bar \Psi}^{L/R}}{\partial t}| \sim | \frac{E'}{\hbar}{\bar \Psi}^{L/R}| \ll \frac{mc^2}{\hbar}|{\bar \Psi}^{L/R}|$, $|\frac{\partial {\bar \Psi}^{L/R}}{\partial x}| \sim |k{\bar \Psi}^{L/R}|=|\frac{p}{\hbar}{\bar \Psi}^{L/R}| \ll \frac{mc}{\hbar}|{\bar \Psi}^{L/R}|$. The limit only works if the potential $A_\mu$ is weak, {\sl i.e.}, $|qA_\mu | \ll mc$ for $\mu = 0, 1$. We define the dimensionless `slow' variables $X = \nu (\frac{mc}{\hbar}) x$ and $T = \nu^2 (\frac{mc^2}{\hbar}) t$ where $\nu$ is a positive real number. The non-relativistic limit is recovered by letting $\nu$ tend to zero while keeping 
$\partial_X {\bar \Psi}^{L/R} = O(1)$, $\partial_T {\bar \Psi}^{L/R} = O(1)$, $\frac{V}{\nu^2} = O(1)$ and $\frac{A_1}{\nu}= O(1)$.

Injecting the above scaling in the KG equation for ${\bar \Psi}_{L/R}$ shows that ${\bar \Psi}_{L/R}$ both obey the Schr\"odinger equation with electromagnetic fields when $\nu$ goes to zero. 
We now also compute for future use the lowest order terms in the difference ${\bar \Psi}_{L} - {\bar \Psi}_{R}$.
The Dirac equation can be rewritten as
\begin{eqnarray}
{\bar \Psi}_{R} & = & {\bar \Psi}_{L} - i \nu D_X {\bar \Psi}_{L} + i \nu^2 D_T {\bar \Psi}_{L} + O(\nu^3),
\nonumber \\ 
{\bar \Psi}_{L} & = & {\bar \Psi}_{R} + i \nu D_X {\bar \Psi}_{R} + i \nu^2 D_T {\bar \Psi}_{R} + O(\nu^3)\label{eq:5}.
\end{eqnarray}

with $D_X=\partial_X+i\frac{qA_1}{mc\nu}$ and $D_T=\partial_T+i\frac{qV}{mc^2\nu^2}$.
 
Using the Schr{\"o}dinger equation to replace the temporal derivatives by spatial derivatives leads to
\begin{eqnarray}
{\bar \Psi}_{R} & = & {\bar \Psi}_{L} - i \nu D_X  {\bar \Psi}_{L} - \frac{\nu^2}{2}D_{XX} {\bar \Psi}_{L} + \mathcal{O}(\nu^3), \nonumber \\ 
{\bar \Psi}_{L} & = & {\bar \Psi}_{R} + i \nu D_X {\bar \Psi}_{R} - \frac{\nu^2}{2}D_{XX} {\bar \Psi}_{R} + \mathcal{O}(\nu^3)\label{eq:9}.
\label{eq:psimeqpsip}
\end{eqnarray}

Let us start the discussion by keeping only the terms of order $\nu$ in Eq.~(\ref{eq:psimeqpsip}). The two wave-function components are equal at order 0 in $\nu$ and thus, at this order, have the same moduli and phases. We want to compute the differences between the moduli and the differences between the phases at first order in $\nu$. This is best done in the following way. 

Write ${\bar \Psi}_{L} = r \exp(\frac{i}{\hbar} \phi)$ and ${\bar \Psi}_{R} = (r + \delta r) \exp\left(\frac{i}{\hbar} (\phi + \delta \phi)\right)$. 
Inserting this into Eq.~(\ref{eq:psimeqpsip}) and keeping only first-order terms leads to: 
\begin{equation}
r \exp (\frac{i}{\hbar} \phi) \left(\frac{i}{\hbar} \delta \phi + \frac{\delta r}{r} \right) = - i \nu D_X {\bar \Psi}_{L},
 \end{equation}
from which one gets:
\begin{equation}
\delta \phi  = - \frac{\hbar}{2 r^2} \nu \left(  {\bar \Psi}_{L}^{*} D_X {\bar \Psi}_{L} + (D_X  {\bar \Psi}_{L})^{*} {\bar \Psi}_{L}\right).
 \end{equation}
The difference $\delta r$ can be obtained in the same manner:
 \begin{equation}
\frac{\delta r}{r}  =  \frac{i}{2 r^2} \nu \left( {\bar \Psi}_{L}^{*} D_X {\bar \Psi}_{L} - (D_X {\bar \Psi}_{L})^{*} {\bar \Psi}_{L}\right).
 \end{equation}
 
This transcribes into:
 \begin{equation}
\delta \phi  = - \nu \frac{\hbar}{r}\, \frac{\partial r}{\partial X},
 \end{equation}
 and
\begin{equation}
\frac{\delta r}{r}  =  \nu \pi .
 \end{equation} 

with $\pi=\frac{1}{\hbar}\frac{\partial \phi}{\partial X}+\frac{qA_1}{mc\nu}$.

It is straightforward (but tedious) to compute in the same manner the differences in moduli and phases at second order in $\nu$. One finds:

 \begin{equation}
\delta \phi  = - \nu \frac{\hbar}{r}\, \frac{\partial r}{\partial X} 
 - \frac{\hbar}{2} \nu^2 \frac{\partial \pi}{\partial X},
 \end{equation}
 and
\begin{align}
\frac{\delta r}{r}  & =  &\nu \pi + \frac{1}{2}\nu^2\left(  \pi^2 \nonumber
 +
\frac{1}{ r^2}\left( \left(\frac{\partial r}{\partial X}\right)^2
- r  \frac{\partial^2 r}{\partial X^2}\right)
\right).
 \end{align} 

\subsubsection*{Hydrodynamical variables}

In the main text the Dirac wavefunction is defined as 
\begin{equation}
 \psi= \frac{1}{\sqrt{2}}e^{\frac{i}{\hbar}\phi_+/2}\begin{pmatrix}
     \sqrt{j^0-j^1}e^{\frac{i}{\hbar}\phi_-/2} \\ \sqrt{j^0+j^1}e^{-\frac{i}{\hbar}\phi_-/2}
    \end{pmatrix}.
\label{def}
\end{equation}
Thus, the  definitions above lead to  $\phi_+=2\phi+\delta \phi -2m c^2 t$ and $\phi_-=-\delta \phi$.

At second order in $\nu$ the hydrodynamical variables defined in the main text read:
\begin{equation}
\begin{split}
n =2 r^2 + 2r^2 \nu \pi
   +\nu^2 r^2 \pi^2 +\nu^2 \left( \left(\frac{\partial r}{\partial X}\right)^2 - r \frac{\partial^2 r}{\partial X^2}\right),
   \label{eq:n}
\end{split}
\end{equation}
\begin{equation}
u^0=c\left(1+ \frac{1}{2}\, \nu^2 \pi^2\right),
\label{eq:u0}
\end{equation}
\begin{equation}
u^1= c\left( \nu  \pi+ \frac{1}{2 r^2}\, \nu^2 \left(\left(\frac{\partial r}{\partial X}\right)^2- r \frac{\partial^2 r}{\partial X^2} \right) \right),
\label{eq:u1}
\end{equation}
\begin{equation}
\begin{split}
w  =  m c^2 \left( 2 r^2+2  r^2 \nu \pi  + \nu^2 \left(
r^2  \pi^2  -    r \frac{ \partial^2 r}{\partial X^2}  \right)\right).
\label{eq:w}
\end{split}
\end{equation}

\subsubsection*{Hydrodynamical equations}

In units where $\hbar \ne 1$ and $c \ne 1$, the relativistic fluid equations can be written as
\begin{equation}
    \frac{1}{c}\partial_{t}(nu^{0})+\partial_{x}(nu^{1})=0,
\end{equation}
\begin{equation}
     \frac{w}{nc}u^{0}=-\frac{\hbar c}{2}\left(\frac{1}{c}\partial_t\phi_++ \partial_{x}(\phi_-)\right)-qV,
\end{equation}
\begin{equation}
     \frac{w}{nc}u^{1}=\frac{\hbar c}{2}\left(\partial_x\phi_++  \frac{1}{c}\partial_{t}(\phi_-)\right)+qA_1c,
\end{equation},
\begin{equation}
     \frac{1}{c}\partial_{t}(nu^{1})+\partial_{x}(nu^{0})=-2\frac{mc^2}{\hbar}n\sin(\phi-).
\end{equation}

By injecting the previous hydrodynamical variables in these equations, one can determine their non-relativistic limit. The set of four independent hydrodynamical relativistic equations become a set of two independent equations in the non-relativistic limit at second order in $\nu$:
\begin{equation}
    \frac{\partial}{\partial T}\left(2r^2\right)+\frac{\partial}{\partial X}\left(2r^2 \pi\right)=0,
\end{equation}
\begin{equation}
        \frac{1}{\hbar}\frac{\partial \phi}{\partial T}+\frac{1}{2}\pi^2+\frac{qV}{mc^2\nu^2}-\frac{1}{2\sqrt{2r^2}}\frac{\partial^2 \sqrt{2r^2}}{\partial X^2}=0.
\end{equation}
Then, in the normal $x$ and $t$ variables one gets the continuity equation and the Bernoulli equation of an inviscid fluid of density $n=2r^2$ and velocity $v=\frac{1}{m}\left(\partial_x\phi+qA_1\right)$ in an electromagnetic potential $V$ and a quantum potential $Q=-\frac{\hbar^2}{2m}\frac{1}{\sqrt{n}}\frac{\partial^2 \sqrt{n}}{\partial x^2}$ called the Bohm potential which vanishes in the non-quantum limit $\hbar\rightarrow0$:
\begin{equation}
    \partial_t(n)+\partial_x(nv)=0,
\end{equation}
\begin{equation}
    \partial_t( \phi)+\frac{1}{2}mv^2+qV+Q=0.
\end{equation}

The gradient of this Bernoulli equation leads to the nonlinear inviscid Burger equation for a charged fluid in an electric field $E=-\partial_xV+\partial_tA_1$  and a quantum pressure force $F_Q=-\partial_xQ$:
\begin{equation}
    m\left(\partial_tv+v\partial_xv\right)=qE+F_Q.
\end{equation}

\end{document}